\def\Title{Computational Complexity of Uniform Quantum Circuit Families
 and Quantum Turing Machines}
\def\Author{Harumichi Nishimura and Masanao Ozawa}
\documentstyle[12pt,mystyle]{article}

  \let\ps=\psi

  \newcommand{\beq}{\begin{equation}}
  \newcommand{\eeq}{\end{equation}}
  \newcommand{\beql}[1]{\begin{equation}\label{eq:#1}}

  \newcommand{\beqa}{\begin{eqnarray}}
  \newcommand{\eeqa}{\end{eqnarray}}
  \newcommand{\beqas}{\begin{eqnarray*}}
  \newcommand{\eeqas}{\end{eqnarray*}}

  \newtheorem{Theorem}{Theorem}[section]
  \newtheorem{Proposition}[Theorem]{Proposition}
  \newtheorem{Lemma}[Theorem]{Lemma}

  \newcommand{\qed}{{\em QED}}

  \newcommand{\C}{{\bf C}}
  
  \newcommand{\N}{{\bf N}}
  \newcommand{\Q}{{\bf Q}}
  \newcommand{\R}{{\bf R}}

  \newcommand{\Z}{{\bf Z}}

  \renewcommand{\l}{\langle}
  \newcommand{\r}{\rangle}

  \newcommand{\bK}{{\bf K}}

  \newcommand{\bN}{{\bf N}}
  
  \newcommand{\bP}{{\bf P}}

  \newcommand{\cB}{{\cal B}}
  \newcommand{\cC}{{\cal C}}
  \newcommand{\cD}{{\cal D}}

  \newcommand{\cG}{{\cal G}}
  \newcommand{\cH}{{\cal H}}

  \newcommand{\cK}{{\cal K}}
  \newcommand{\cL}{{\cal L}}

  \newcommand{\cR}{{\cal R}}

  \newcommand{\cX}{{\cal X}}

  \newcommand{\rP}{{\rm P}}

  \newcommand{\al}{\alpha}                                             %
  \newcommand{\da}{\dagger}
  \newcommand{\de}{\delta}                                             %
  \newcommand{\ep}{\epsilon}
  \newcommand{\et}{\eta}
  \newcommand{\ga}{\gamma}                                             %

  \newcommand{\la}{\lambda}                                            %
  \newcommand{\mb}{\mbox}
  
  \newcommand{\ph}{\phi}                                               %
  \newcommand{\rh}{\rho}
  \newcommand{\si}{\sigma}                                             %
  \newcommand{\ta}{\tau}                                               %
  \newcommand{\th}{\theta}                                             %
  \newcommand{\La}{\Lambda}

  \newcommand{\Si}{\Sigma}

  \newcommand{\beqan}{\begin{eqnarray*}}
  \newcommand{\beqar}[1]{\begin{equation}\label{#1}\begin{array}{l}}

  \newcommand{\eeqar}{\end{array}\end{equation}}
 \renewcommand{\ep}{\varepsilon}

\newcommand{\bra}[1]{\langle#1|}
\newcommand{\ket}[1]{|#1\rangle}
\newcommand{\braket}[1]{\langle#1\rangle}

\newcommand{\bold}[1]{\mbox{\boldmath{$#1$}}}
  \title{\bf \Title}
  \author{\sc \Author \\ 
  \small\em  Graduate School of Human Informatics and
 School of Informatics and Sciences\\
  \small\em  Nagoya University, Nagoya 464-8601, Japan}
  \date{}
\begin{document}
\maketitle
\begin{abstract}
 Deutsch proposed two sorts of models of quantum computers,
 quantum Turing machines (QTMs) and quantum circuit families (QCFs). 
 In this paper we explore the computational powers of
 these models and re-examine the claim of
 the computational equivalence of these models often made
 in the literature without detailed investigations.
 For this purpose, we formulate the notion of the codes
 of QCFs and the uniformity of QCFs by the computability
 of the codes. Various complexity classes are introduced
 for QTMs and QCFs according to
 constraints on the error probability of algorithms
 or transition amplitudes.
 Their interrelations are examined in detail.
 For Monte Carlo algorithms, it is proved that
 the complexity classes based on uniform QCFs
 are identical with the corresponding classes based on QTMs.
 However, for Las Vegas algorithms, it is still open
 whether the two models are equivalent.
 We indicate the possibility that they are not equivalent.
 In addition, we give a complete proof of the existence
 of a universal QTM simulating multi-tape QTMs efficiently.
 We also examine the simulation
 of various types of QTMs such as multi-tape QTMs, single tape QTMs,
 stationary, normal form QTMs (SNQTMs), and
 QTMs with the binary tapes. As a result,
 we show that these QTMs are computationally equivalent
 one another as computing models implementing not only
 Monte Carlo algorithms but exact (or error-free) ones.  
\end{abstract}


\section{Introduction}

In the early 1980's, Feynman \cite{Fey82} suggested that
 computers based on quantum mechanics
 would carry out computations more efficiently than classical ones,
 and Benioff \cite{Ben82} started the study of
 quantum mechanical Hamiltonian models of Turing machines.
 In the late 1980's, Deutsch introduced
 quantum Turing machines (QTMs) \cite{Deu85}
 and quantum circuits \cite{Deu89} as models of quantum computers. 
 Using Deutsch's models, several results were obtained
 to suggest that quantum computers are more powerful than
 classical ones \cite{BB94,BV93,DJ92,Sim94}.
 Eventually, Shor \cite{Sho94} found efficient quantum algorithms
 for the factoring problem and the discrete logarithm problem,
 which are considered to have no efficient algorithms
 in computational complexity theory and applied to
 public-key cryptosystems.
 Since then, many experiments have been attempted
 to realize a quantum computer. 

 Up to now, the above two models
 appear to be studied under different objectives.
 A QTM models a programmable computing machine
 and has been used as a mathematical model for studying
 the efficiency of quantum computation.
 On the other hand,
 a quantum circuit has been studied mainly as a physical model
 for realization. Thus, in order to make a bridge
 between these two approaches, it is
 important to give a detailed comparison of their computational powers
 from a complexity theoretical point of view. 

 The existence of a universal QTM was shown first
 by Deutsch \cite{Deu85}.
 However, his universal QTM needs exponential slowdown
 for simulating QTMs.
 In 1993, Bernstein and Vazirani \cite{BV93}
 claimed that there is an efficient universal QTM,
 and gave a detailed proof in \cite{BV97}.
 But their universal QTM is applicable only to
 QTMs such that the head must move either to the right
 or to the left at each step (two-way QTMs),
 and their method cannot afford
 an efficient simulation of a general QTM
 with its head allowed not to move.
 Shortly after, Yao \cite{Yao93} claimed
 the existence of a universal QTM simulating general QTMs,
 with the following sketch of the proof:
 He first shows that there is a quantum circuit
 simulating a given QTM for arbitrary steps,
 and his universal QTM is designed to carry out
 each step of the computing of the quantum circuit.
 This sketch also contains the existence of a QTM 
 that simulates any quantum circuit implicitly.
 From the above argument, it is often claimed in the literature
 that quantum circuits and QTMs are computationally equivalent. 
 However, from the computational complexity theoretical point of view,
 the following points are left for further investigations.

 In the first place,
 Yao did not define the uniformity of quantum circuit families (QCFs).
 Since a single quantum circuit has a constant input length,
 we need to consider families of quantum circuits for
 comparing the computational power of quantum circuits with QTMs.
 From the viewpoint of polynomial complexity, 
 it is well known that Boolean circuit families with arbitrary input length
 should satisfy a uniformity condition,
 as long as they are computationally no more powerful than
 Turing machines.
 The uniformity of QCFs was mentioned shortly
 by Ekert-Jozsa \cite{EJ96} and Shor \cite{Sho97}.
 As pointed out by Shor,
 we need to introduce a definition of uniformity
 quite different from Boolean circuit families,
 because each wire has continuously many different states
 rather than only two in Boolean circuits.
 Secondly, the complexity classes of QCFs have not been defined explicitly.
 Shor \cite{Sho97} claimed that QTMs and QCFs are equivalent
 as probabilistic computing models implementing
 Monte Carlo algorithms, but the proof has not been given.
 Moreover, it has not been discussed yet
 whether two models are equivalent as probabilistic machines
 implementing Las Vegas algorithms
 or exact algorithms (algorithms which always produce correct answers).
 In order to study these problems, we should set up
 various complexity classes for QTMs and QCFs
 according to constraints on the algorithms.

In this paper, we shall introduce the rigorous formulation
 of uniformity of QCFs
 and investigate the detailed relationship among
 complexity classes of QTMs and uniform QCFs.
 We introduce the class {\bf BUPQC} of languages
 that are efficiently recognized by Monte Carlo type uniform QCFs,
 and show that {\bf BUPQC} coincides with the class
 {\bf BQP} of languages that are efficiently recognized
 by Monte Carlo type QTMs.
 On the other hand, we show that the class {\bf ZQP} of languages that 
 are efficiently recognized by Las Vegas type QTMs is included
 in the class {\bf ZUPQC} that are efficiently recognized
 by Las Vegas type uniform QCFs. However, it remains still open 
 whether these models are equivalent as
 computing models implementing Las Vegas algorithms. 
 Moreover,
 we indicate the possibility that the inclusion is proper.

 In addition, we discuss the relationship among various types of
 QTMs, in particular, single tape QTMs and multi-tape QTMs.
 In the classical case, it is possible to simulate
 a multi-tape Turing machine by a single tape Turing machine
 with quadratic polynomial slowdown.
 Multi-tape QTMs are indispensable
 to examine the $o(n)$-space bounded complexity
 or count the number of steps of a QTM. Thus
 it is important to investigate the level of
 the computational equivalence
 of single tape QTMs and multi-tape QTMs.
 
 We generalize Yao's construction of quantum circuits
 simulating single tape QTMs to multi-tape QTMs and
 give a complete proof of the existence of
 a single tape universal QTM simulating multi-tape QTMs efficiently.
 This shows that a multi-tape QTM can be simulated
 with arbitrary accuracy by a single tape QTM
 with polynomial slowdown.
 We also examine the simulation of various types of QTMs
 such as multi-tape QTMs, single tape QTMs,
 stationary, normal form QTMs (SNQTMs), and
 QTMs with the binary tapes.
 As a result, we show that these QTMs are
 computationally equivalent one another as computing models
 implementing not only Monte Carlo algorithms but exact ones.  

This paper is organized as follows.
 In section 2 we give definitions on QTMs and explain related notions.
 In section 3, we adapt some of basic lemmas on QTMs given
 by Bernstein and Vazirani \cite{BV97} to the present approach.
 Moreover, we show that QTMs with the binary tapes
 are equivalent to two-way QTMs as computing models
 implementing exact algorithms.   
 In section 4 we show that there is a universal QTM
 simulating multi-tape QTMs.
 This section also contains the rigorous formulation of quantum circuits.
 In section 5 we formulate the uniformity of QCFs
 and introduce various classes of languages recognized
 by QTMs and uniform QCFs.
 We also show that QTMs and uniform QCFs are equivalent
 as probabilistic computing models implementing Monte Carlo algorithms,
 we indicate the possibility
 that these two models are not computationally equivalent
 as computing models implementing Las Vegas algorithms,
 and we show that SNQTMs are equivalent
 to multi-tape QTMs as computing models implementing exact algorithms. 

\section{Quantum Turing Machines}

In what follows, for any integers $n<m$ the interval 
$\{n,n+1,\ldots,m-1,m\}$ is denoted by $[n,m]_\Z$.
 A quantum Turing machine (QTM) $M$ is a quantum system
 consisting of a processor, a bilateral infinite tape
 and a head to read and write a symbol on the tape.
 We refer to Deutsch \cite{Deu85} for the physical formulation of a QTM.
 The formal definition of a QTM as a mathematical structure 
 is given as follows. 
A {\em processor configuration set} is a finite set with two specific
 elements denoted by $q_0$ and $q_f$, where 
 $q_0$ represents the {\it initial processor configuration}
 and $q_f$ represents the {\it final processor configuration}.
 A {\it symbol set} is a finite set of cardinality at least 2
 with a specific element denoted by $B$ and called the {\it blank}. 
 A {\it tape configuration} from a symbol set $\Si$ is a function $T$ from
 the set $\Z$ of integers to $\Si$ such that $T(m)=B$ except for
 finitely many $m\in\Z$. The set of all the possible tape configurations
 is denoted by $\Si^{\#}$. The set $\Si^{\#}$ is a countable set.
 For any $T\in\Si^\#$, $\ta\in\Si$, and $\xi\in\Z$, the tape configuration
 $T_\xi^{\ta}$ is defined by 
$$ 
T_\xi^{\ta}(m)=\left\{
\begin{array}{ll}
\ta &\mbox{if}\ m=\xi,\\
T(m)&\mbox{if}\ m\neq \xi.
\end{array}\right.
$$
 A {\em Turing frame} is a pair $(Q,\Si)$ of
 a processor configuration set $Q$ and a symbol set $\Si$.
 In what follows, let $(Q,\Si)$ be a Turing frame.
 The {\it configuration space} of $(Q,\Si)$ is the product set
 $\cC(Q,\Si)=Q\times\Si^{\#}\times\Z$.
 A {\it configuration} of $(Q,\Si)$ is an element $C=(q,T,\xi)$
 of $\cC(Q,\Si)$. Specifically, if $q=q_0$ and $\xi=0$ then $C$ is called 
an {\it initial configuration} of $(Q,\Si)$, 
and if $q=q_f$ then $C$ is called a {\it final configuration} of $(Q,\Si)$.
 The {\em quantum state space} of $(Q,\Si)$ is the Hilbert space 
$\cH(Q,\Si)$ spanned by $\cC(Q,\Si)$ with the canonical basis  
$\{\ket{C}|\ C\in\cC(Q,\Si)\}$ called the {\em computational basis}.
 A {\it quantum transition function} for $(Q,\Si)$ 
is a function from $Q\times\Si\times Q\times\Si\times[-1,1]_\Z$ 
into the complex number field $\C$.
 A {\em (single tape) prequantum Turing machine} is
 defined to be a triple $M=(Q,\Si,\de)$ consisting of a Turing frame
 $(Q,\Si)$ and a quantum transition function $\de$ for $(Q,\Si)$.

 Let $M=(Q,\Si,\de)$ be a prequantum Turing machine.  
 An element of $Q$ is called a {\it processor configuration} of $M$, 
the set $\Si$ is called the {\it alphabet} of $M$, 
the function $\de$ is called the {\it quantum transition function} of $M$,
 and an (initial or final) configuration of $(Q,\Si)$ is called
 an {\em (initial or final) configuration} of $M$.
 A unit vector in $\cH(Q,\Si)$ is called a {\em state} of $M$.
 The {\it evolution operator} of $M$ is
 a linear operator $M_\de$ on $\cH(Q,\Si)$ such that 
$$ 
M_\de\ket{q,T,\xi}=\sum_{p\in Q,\ta\in\Si,d\in[-1,1]_\Z}
\de(q,T(\xi),p,\ta,d)\ket{p,T_\xi^{\ta},\xi+d}
$$
for all $(q,T,\xi)\in\cC(Q,\Si)$. 
The above equation uniquely defines the bounded operator $M_\delta$ 
on the space $\cH(Q,\Si)$.
 A (single tape) prequantum Turing machine
 is said to be a {\em (single tape) quantum Turing machine (QTM)}
 if the evolution operator is unitary.

 A quantum transition function $\de$ for $(Q,\Si)$ is said
 to be {\em two-way} 
 if $\de(p,\si,q,\ta,0)=0$ for any $(p,\si,q,\ta)\in(Q\times\Si)^2$.
 A prequantum Turing machine (or QTM) $M=(Q,\Si,\de)$ is said to be two-way 
 if $\de$ is two-way (In \cite{BV97}, two-way QTMs are merely called QTMs, 
 and QTMs in this paper are called general QTMs). 

 The following theorem proved in \cite{ON98} characterizes
 the quantum transition functions that give rise to QTMs.
 The quantum transition function of a two-way QTM
 satisfies condition (c) of Theorem \ref{th:21} automatically.
 In this case, Theorem \ref{th:21} is reduced to the result
 due to Bernstein and Vazirani \cite{BV93,BV97}.

\begin{Theorem}\label{th:21}
 A prequantum Turing machine $M=(Q,\Si,\de)$ is a QTM 
if and only if $\de$ satisfies the following conditions.

{\rm (a)} For any $(q,\si)\in Q\times\Si$,
$$ \sum_{p\in Q,\ta\in\Si,d\in[-1,1]_\Z}|\de(q,\si,p,\ta,d)|^2=1.$$

{\rm (b)} For any $(q,\si),(q',\si')\in Q\times \Si$ with $(q,\si)\neq
(q',\si')$,
$$ \sum_{p\in Q,\ta\in\Si,d\in[-1,1]_\Z}\de(q',\si',p,\ta,d)^{*}
\de(q,\si,p,\ta,d)=0. 
$$

{\rm (c)} For any $(q,\si,\ta),(q',\si',\ta')\in Q\times \Si^2$,
$$ \sum_{p\in Q,d=0,1}\de(q',\si',p,\ta',d-1)^{*}\de(q,\si,p,\ta,d)
=0. $$

{\rm (d)} For any $(q,\si,\ta),(q',\si',\ta')\in Q\times \Si^2$,
$$ \sum_{p\in Q}\de(q',\si',p,\ta',-1)^*\de(q,\si,p,\ta,1)=0.$$
\end{Theorem}

Ko and Friedman \cite{KF82} introduced
 the notion of efficiently computable numbers.
 A real number $x$ is {\em polynomial time computable}
 if there is a polynomial time computable function $\ph$
 such that $|\ph(1^n)-x|\le 2^{-n}$ and $\ph(1^n)\in \{m/2^n|\ m\in\Z\}$ 
 for any $n\in \N$. We denote by $\rP\R$ the set of polynomial time
 computable real numbers and let $\rP\C=\{x+y\sqrt{-1}|x,y\in \rP\R\}$.
 We say that a QTM $M=(Q,\Si,\de)$ is {\em in $\rP\C$}
 if the range of $\de$ is included in $\rP\C$.
 In this paper, we define a QTM to be with amplitudes in $\C$,
 since in section 5 we investigate QTMs with amplitudes in $\C$
 as a mathematical object. However, 
 from the complexity theoretical point of view,
 we need to require that QTMs are in $\rP\C$
 as defined by Bernstein and Vazirani \cite{BV97}.
 When we consider a universal QTM in section 4,
 we also restrict the QTMs given as the input of
 the universal QTM to QTMs in $\rP\C$,
 since not every QTM can be (efficiently) encoded
 with absolute accuracy by classical means.
 We now define the {\em code $c(x)$ of an element $x$ in $\rP\R$}
 by the code of a polynomial time bounded
 deterministic Turing machine
 computing one of its rational approximations,
 and define the code of an element $z=x+y\sqrt{-1}$ in $\rP\C$
 by $c(z)=\l c(x),c(y)\r$.
 Then the QTMs in $\rP\C$ can be easily encoded:
 we define the {\em code of a QTM} $M=(Q,\Si,\de)$ in $\rP\C$
 to be the list of the codes of elements
 $\de(q,\si,p,\ta,d)$ in $\rP\C$,
 where $(q,\si,p,\ta,d)\in Q\times\Si\times Q
\times\Si\times [-1,1]_\Z$.

 A finite string from a symbol set $\Si$ is called a $\Si$-string.
 The length of a $\Si$-string $x$ is denoted by $|x|$ and
 the set of all the possible $\Si$-strings is denoted by $\Si^*$.
 A tape configuration $T$ from $\Si$ is said to {\em represent}
 a $\Si$-string $x=\si_0\cdots \si_{k-1}$ of length $k$, if $T$ satisfies
$$ 
T(m)=\left\{
\begin{array}{ll}
\si_m&\mb{if}\ m\in[0,k-1]_\Z,\\
B&\mb{otherwise}.
\end{array}\right.
$$
 In what follows, we denote by ${\rm tape}[x]$ the tape configuration
 representing $x$.

 For symbol sets $\Si_1,\ldots,\Si_k$ with the blanks $B_1,\ldots,B_k$, 
 the product set $\Si=\Si_1\times\cdots\times\Si_k$ can be considered 
 as a symbol set with the blank $B=(B_1,\ldots,B_k)$.
 The projection from $\Si$ to $\Si_i$ is denoted by $\pi_i$. 
 If $s_i=\si_{1i}\cdots\si_{ni}$ is a $\Si_i$-string of length $n$
 for $i=1,\ldots,k$,
 the $\Si$-string $(\si_{11},\ldots,\si_{1k})\cdots 
(\si_{n1},\ldots,\si_{nk})$ is also denoted by $(s_1,\ldots,s_k)$.
 A {\it $k$-track} QTM is such that the alphabet $\Si$ is factorized as 
 $\Si=\Si_1\times\cdots\times\Si_k$ with symbol sets $\Si_1,\ldots,\Si_k$. 
 The symbol set $\Si_i$ is called the $i$-th track alphabet of this QTM.
 If the tape configuration is $T$, the $i$-th track configuration
 is defined as the function $T^i=\pi_iT\in\Si_i^\#$,
 so that we have $T(m)=(T^1(m),\ldots,T^k(m))$ for any $m\in\Z$.
 For $i=1,\ldots,j$, let $s_i$ be a $\Si_i$-string
 of length at most $n$ and $s_iB^{n_i}$
 be the $\Si_i$-string $s_iBB\cdots B$ of length $n$.
 Then, ${\rm tape}[s_1,\ldots,s_j]$ abbreviates
 ${\rm tape}[(s_1B^{n_1},\ldots,s_jB^{n_j},
 \underbrace{B^{n},\ldots,B^{n}}_{k-j})]$.

Let a symbol set $\Si$ be decomposed as $\Si=\Si_1\times\cdots\times\Si_k$.
 The quantum state space $\cH(Q,\Si)$ can be factorized as 
 $\cH(Q,\Si)= \cH(Q)\otimes\cH(\Si^{\#})\otimes\cH(\Z)$ or  
 $\cH(Q,\Si)= \cH(Q)\otimes\cH(\Si_1^{\#})
\otimes\cdots\otimes\cH(\Si_k^{\#})\otimes\cH(\Z),$
 where $\cH(Q),\ \cH(\Si^{\#}),\ \cH(\Si_i^{\#}),$ and $\cH(\Z)$ are
 the Hilbert spaces generated by $Q,\ \Si^{\#},\ \Si_i^{\#}$ and $\Z$,
 respectively.
 Then, the computational basis state $\ket{q,T,\xi}$ can be represented
 as $\ket{q,T,\xi}=\ket{q}\ket{T}\ket{\xi}$ or
 $\ket{q,T,\xi}=\ket{q}\ket{T^1}\cdots\ket{T^k}\ket{\xi}$ 
 by the canonical bases $\{\ket{q}|\ q\in Q\}$ of $\cH(Q)$, 
 $\{\ket{T}|\ T\in\Si^{\#}\}$ of $\cH(\Si^{\#})$,
 $\{\ket{T^i}|\ T^i\in\Si_i^{\#}\}$ of $\cH(\Si_i^{\#})$, and
 $\{\ket{\xi}|\ \xi\in\Z\}$ of $\cH(\Z)$.
 
 Let $M=(Q,\Si,\de)$ be a QTM, and we assume the numbering of 
 $Q$ and $\Si$ such that $Q=\{q_0,\ldots,q_{|Q|-1}\}$ 
 and $\Si=\{\si_0,\ldots,\si_{|\Si|-1}\}$, 
 where we denote by $|X|$ the cardinality of a set $X$. 
 We define projections $E^{\hat{q}}(q_j)$,
 $E^{\hat{T}(m)}(\si_j)$ for $m\in\Z$, 
 and $E^{\hat{\xi}}(\xi)$ for $\xi\in\Z$ by
$$
E^{\hat{q}}(q_j)=\ket{q_j}\bra{q_j}\otimes I_2\otimes I_3,\ \ \ 
E^{\hat{T}(m)}(\si_j)=
\sum_{T(m)=\si_j}I_1\otimes |T\r\l T|\otimes I_3,\ \ \ 
E^{\hat{\xi}}(\xi)=I_1\otimes I_2\otimes\ket{\xi}\bra{\xi},
$$ 
 where $I_1$, $I_2$, and $I_3$ are the identity operators 
 on $\cH(Q),\cH(\Si^{\#})$, and $\cH(\Z)$, respectively. 
 Moreover, if $M$ is a $k$-track QTM with
 alphabet $\Si=\Si_1\times\cdots\times\Si_k$,
 we define a projection $E^{\hat{T^i}}(T^i)$ for $T^i\in\Si_i^\#$
 where $i=1,\ldots,k$ by
$$ 
 E^{\hat{T^i}}(T^i)=
I_1 \otimes I_{2,1}\otimes\cdots\otimes I_{2,i-1}\otimes 
|T^i\r\l T^i|\otimes I_{2,i+1}\otimes\cdots\otimes I_{2,k}\otimes I_3,
$$
 where $I_{2,j}$ is the identity operator on $\cH(\Si_j^{\#})$. 

 A QTM $M=(Q,\Si,\de)$ is said to be {\it stationary}
 \cite[Definition 3.12]{BV97},
 if for every initial configuration $C$,
 there exists some $t\in\N$ such that
 $||E^{\hat{\xi}}(0)E^{\hat{q}}(q_f)M_\de^{t}\ket{C}||^2=1$ and
 for all $s<t$ we have $||E^{\hat{q}}(q_f)M_\de^{s}\ket{C}||^2=0$.
 The positive integer $t$ is called the {\it computation time} of $M$
 for input state $\ket{C}$.
 Specifically, if $\ket{C}=\ket{q_0,{\rm tape}[x],0}$,
 it is called the computation time of $M$ on input $x$. 
 A {\it polynomial time bounded} QTM is a stationary QTM such that 
 on every input $x$ the computation time is bounded by
 a polynomial in the length of $x$.
 Moreover, let $\ket{\ph}=\sum_{x\in\Si^n}\al_x\ket{q_0,{\rm tape}[x],0}$
 for some $n\in\N$.
 Then, if the computation time of $M$ on every input $x$
 satisfying $\al_x\neq 0$ is $t$, 
 the state $M_\de^t\ket{\ph}$ is called the {\it output state} of $M$
 for {\it input state} $\ket{\ph}$. 
 A QTM $M=(Q,\Si,\de)$ is said to be in {\it normal form}
 \cite[Definition 3.13]{BV97},
 if $\de(q_f,\si,q_0,\si,1)=1$ for any $\si\in\Si$.
 In what follows ``SNQTM'' abbreviates ``stationary, normal form QTM''.
 We may consider only SNQTMs without loss of generality to develop
 quantum complexity theory as shown later (Theorem \ref{th:57}).

 Finally, we shall give a formal definition of simulation.
 Let $M=(Q,\Si,\de)$ and $M'=(Q',\Si',\de')$ be QTMs.
 Let $t$ be a positive integer and $\ep>0$. 
 Let $e:\cC(Q,\Si)\rightarrow \cC(Q',\Si')$ be an injection computable
 in polynomial time, $d:\cC(Q',\Si')\rightarrow \cC(Q,\Si)$ a function
 computable in polynomial time satisfying $d\cdot e=\mb{id}$,
 and $f$ a function from $\N^2$ to $\N$.
 We say that $M'$ {\it simulates} $M$ for $t$ steps with {\it accuracy} $\ep$
 and {\it slowdown} $f$ (under the {\it encoding} $e$ and the {\it decoding}
 $d$), if for any $C_0\in\cC(Q,\Si)$, we have
\begin{equation}\label{eq:21}
 \sum_{C'\in\cC(Q,\Si)}\left|\left|\l C'|M_\de^t|C_0\r\right|^2-
 \sum_{C\in d^{-1}(C')}\left|\l C|{M}^{f(t,\lceil\frac{1}{\ep}\rceil)}_{\de'}|
 e(C_0)\r\right|^2\right|\le\ep. 
\end{equation}
 If $f$ depends only on $t$ and Eq.~(\ref{eq:21}) is satisfied
 for $\ep=0$, we merely say that $M'$ simulates $M$ for $t$ steps
 with slowdown $f$.
 In particular, we say that $M'$ simulates $M$ for $t$ steps
 {\it by a factor of} $s$ if $f(t)=st$.

We have discussed solely single tape QTMs,
 but our arguments can be adapted easily to multi-tape QTMs. 
 We refer to \cite{ON98} for the formulation of multi-tape QTMs.
  
\section{Basic Lemmas for QTMs}

In this section, we present several definitions, lemmas and
 theorems necessary to prove theorems in sections 4 and 5. 
 Except for Lemma \ref{th:32}, they are given by
 Bernstein and Vazirani \cite{BV97}
 and we adapt them to the present approach.
 We refer to \cite{BV97} for these proofs.
 In \cite{BV97}, the dovetailing lemma and the branching lemma are given
 for two-way QTMs, but we extend them to general QTMs
 including multi-tape QTMs. 

Let $S\subseteq Q\times\Si$.
 A complex-valued function $\de$ on $S\times Q\times\Si\times[-1,1]_\Z$
 is {\it unidirectional},
 if we have $d=d'$ whenever $\de(p,\si,q,\ta,d)$ and
 $\de(p',\si',q,\ta',d')$ are both non-zero,
 where $q\in Q$, $(p,\si),(p',\si')\in S$, $\ta,\ta'\in\Si$,
 and $d,d'\in[-1,1]_\Z$.
 A prequantum Turing machine (or QTM) 
 is said to be {\it unidirectional}
 if the quantum transition function is unidirectional.
 This definition is a natural extension of the definition
 of \cite{BV97} to the case where the head is not required to move.
 It is easy to see that a unidirectional
 prequantum Turing machine is a unidirectional QTM
 if it satisfies conditions (a) and (b) of
 Theorem \ref{th:21}. We can show the following lemma
 for a unidirectional QTM by the similar way to \cite{BV97}. 
 This lemma allows us to extend a partially defined 
 unidirectional quantum transition function to characterize a QTM.

\begin{Lemma}[Completion lemma]\label{th:31} 
 Let $\de'$ be a unidirectional function
 on $S\times Q\times\Si\times[-1,1]_\Z$,
 where $S\subseteq Q\times\Si$.
 Assume that $\de'$ satisfies
 the following conditions {\rm (a)} and {\rm (b)},

{\rm (a)} For any $(q,\si)\in S$,
$$ 
\sum_{p\in Q,\ta\in\Si,d\in [-1,1]_\Z}|\de'(q,\si,p,\ta,d)|^2=1.
$$

{\rm (b)} For any $(q,\si),(q',\si')\in S$ with $(q,\si)\neq
(q',\si')$,
$$ 
\sum_{p\in Q,\ta\in\Si,d\in [-1,1]_\Z}
\de'(q',\si',p,\ta,d)^*\de'(q,\si,p,\ta,d)=0. 
$$
 Then there is a unidirectional QTM $M=(Q,\Si,\de)$ such that 
 $\de(p,\si,q,\ta,d)=\de'(p,\si,q,\ta,d)$
 whenever $\de'(p,\si,q,\ta,d)$ is defined.
\end{Lemma}

As is well-known, any deterministic Turing machine (DTM) $M=(Q,\Si,\de)$
 can be simulated by a DTM $M'=(Q',\{B,1\},\de')$ with slowdown
 by a factor of $\lceil\log |\Si|\rceil$.
 Using the completion lemma, we can prove a similar statement
 for unidirectional QTMs.
 
\begin{Lemma}\label{th:32}
 Any unidirectional QTM $M=(Q,\Si,\de)$ 
 can be simulated by a unidirectional QTM $M'=(Q',\{B,1\},\de')$
 with slowdown by a factor of $3k$, where $k=\lceil\log |\Si|\rceil$. 
\end{Lemma}

{\it Proof.}\ Throughout this proof, 
 we denote by $\si_0\cdots\si_{k-1}$ the binary representation of $\si\in\Si$.
 Let $Q'=(Q\times\{1\})\cup (\bigcup_{j=1}^k(Q\times\{B,1\}^j\times\{1,2\}))
  \cup (Q\times [1,k-1]_\Z\times\{3\})$. 
 We define the function $e:\cC(Q,\Si)\rightarrow\cC(Q',\{B,1\})$
 such that $e(p,T,\xi)=(p,\widetilde{T},k\xi)$,
 where $\widetilde{T}$ is the tape configuration from $\{B,1\}$ such that  
 $\widetilde{T}(kj)\cdots\widetilde{T}(kj+k-1)=\si_0\cdots\si_{k-1}$
 if $T(j)=\si$ for any $j\in\Z$, that is, the function $e$ determines
 the configuration of $M'$ corresponding to a configuration of $M$.
 If a state $\ket{p,T,\xi}$ of $M$ such that $T(\xi)=\si$
 evolves to $|q,T_\xi^\ta,\xi+d\r$ with amplitude $\de(p,\si,q,\ta,d)$, 
 the corresponding state $|p,\widetilde{T},k\xi\r$ of $M'$ 
 evolves to $|q,\widetilde{T}_\xi^{(\ta)},k(\xi+d)\r$
 with the same amplitude in $3k$ steps
 by the following function $\de'$ on
 $S=(Q_1\times\{B,1\})\cup (Q_2\times\{B\})$.
\begin{eqnarray}
& &\label{eq:31}\!\!\!\!\!\!\de'((p,\si_0,\ldots,\si_{i-1},1),\si_i,
(p,\si_0,\ldots,\si_i,1),B,1)=1\ \ \ \ \ (0\le i\le k-1)\\
& &\label{eq:32}\!\!\!\!\!\!\de'((p,\si_0,\ldots,\si_{k-1},1),b,
(q,\ta_0,\ldots,\ta_{k-1},2),b,-1)=\de(p,\si,q,\ta,d)\ \ \ (b\in\{B,1\})\\
& &\label{eq:33}\!\!\!\!\!\!\de'((q,\ta_0,\ldots,\ta_i,2),B,
(q,\ta_0,\ldots,\ta_{i-1},2),\ta_i,-1)=1\ \ \ \ (1\le i\le k-1)\\
& &\label{eq:34}\!\!\!\!\!\!\de'((q,\ta_0,2),B,(q,1,3),\ta_0,d)=1\\
& &\label{eq:35}\!\!\!\!\!\!\de'((q,i,3),\ta_i,(q,i+1,3),\ta_i,d)=1\ \ \ \ \
 (1\le i\le k-1,\ (q,k,3)=(q,1)).
\end{eqnarray}
 Here we put $Q_1=(\bigcup_{j=0}^{k-1}(Q\times\{B,1\}^j\times\{1\}))\cup
(Q\times\Si'\times\{1\})\cup (Q\times [1,k-1]_\Z\times\{3\})$,
 where $\Si'$ is the subset of $\{0,1\}^k$ corresponding to $\Si$,
 we put $Q_2=\bigcup_{j=1}^k (Q\times\{B,1\}^j\times\{2\})$,
 and $\widetilde{T}^{(\ta)}_\xi$ is
 the tape configuration from $\{B,1\}$ defined by
$$
\widetilde{T}^{(\ta)}_\xi(m)=\left\{
\begin{array}{ll}
\ta_{m\ {\rm mod}\ k} &\mbox{if}\ k\xi\le m\le k\xi+k-1,\\
\widetilde{T}(m)      &\mbox{otherwise}.
\end{array}\right.
$$
 For any element $(p,\si,q,\ta,d)$ except the elements
 defined by the above equations, we define $\de'(p,\si,q,\ta,d)=0$. 
 Eq.~(\ref{eq:31}) represents the operation of recording
 the current symbol $\si$ scanned by the head of $M$ 
 in the processor of $M'$ in $k$ steps.
 Eq.~(\ref{eq:32}) represents the operation of transforming
 the processor configuration $p$
 and the symbol $\si$ of $M$ recorded in the processor of $M'$
 to a new processor configuration $q$ and symbol $\ta$
 with amplitude $\de(p,\si,q,\ta,d)$.
 Since $M$ is unidirectional,
 the direction $d$ in which the head of $M$ moves is uniquely
 determined by $q$.
 Eqs.~(\ref{eq:33}) and (\ref{eq:34}) represent the operation of writing
 the symbol string corresponding to the new symbol $\ta$ of $M$
 in turn on $k$ cells of $M'$ in $k$ steps.
 Eq.~(\ref{eq:35}) represents the operation of moving the head of $M'$
 to the direction $d$ in $k-1$ steps.
 By the above operations, $M'$ carries out the operation corresponding to
 one step of $M$.

 We can see that the function $\de'$ is unidirectional and
 satisfies conditions (a) and (b) of the completion lemma,
 so that there exists a quantum transition function 
 that carries out the above steps by the completion lemma.
 It is easy to see that $M'$ simulates $M$ with slowdown 
 by a factor of $3k$.\ \ \ \qed 

Since every two-way QTM is simulated by a unidirectional QTM 
 with slowdown by a factor of 5 \cite[Lemma 5.5]{BV97},
 Lemma \ref{th:32} implies that any two-way QTM is simulated
 by a unidirectional QTM with the binary tape
 with slowdown by a constant factor independent of the input.

A reversible Turing machine (RTM) $M=(Q,\Si,\de)$
 with classical transition function $\de$
 can be canonically identified with the QTM $M'=(Q,\Si,\de')$ such that 
 the range of $\de'$ is $\{0,1\}$ and that $\de'(p,\si,q,\ta,d)=1$ 
 if and only if $\de(p,\si)=(q,\ta,d)$
 for any $(p,\si,q,\ta,d)\in Q\times\Si\times Q\times\Si\times[-1,1]_\Z$.
 We consider that the class of RTMs is a subclass of the class of QTMs
 under this identification.
 Then, the RTM identified with an SNQTM is called
 a stationary, normal form RTM and we abbreviate it as an ``SNRTM''.
 
\begin{Theorem}[Synchronization theorem]\label{th:33} 
 If $f$ is a function mapping symbol strings to symbol strings
 which can be computed by a DTM in polynomial time
 and if $|f(x)|$ depends only on $|x|$,
 then there is a two-way SNRTM such that
 the output state for input state $\ket{q_0,{\rm tape}[x],0}$
 is $\ket{q_f,{\rm tape}[x,f(x)],0}$
 and whose computation time is a polynomial in $|x|$.
 Moreover, if $f$ and $f^{-1}$ can be computed by
 DTMs in polynomial time and if $|f(x)|$ depends only on $|x|$,
 then there is a two-way SNRTM
 such that the output state for input state $\ket{q_0,{\rm tape}[x],0}$
 is $\ket{q_f,{\rm tape}[f(x)],0}$
 and that the computation time is a polynomial in $|x|$.
\end{Theorem}

 Given any QTM $M=(Q,\Si,\de)$ and any symbol set $\Si'$,
 the QTM $M(\Si')=(Q,\Si\times\Si',\de')$ is
 called the QTM constructed by the {\it addition} of the track
 (with alphabet $\Si'$ to $M$) if
 for any $(p,(\si,\si'),q,(\ta,\ta'),d)\in (Q\times
 (\Si\times\Si'))^2\times[-1,1]_\Z$, we have
$$
\de'(p,(\si,\si'),q,(\ta,\ta'),d)=
\bold{\de}_{\si'}^{\ta'}\de(p,\si,q,\ta,d),
$$ 
 where $\bold{\de}$ denotes the Kronecker delta.
 Given any $k$-track QTM $M=(Q,\Si_1\times \cdots\times \Si_k,\de)$
 and any permutation $\pi:[1,k]_\Z\rightarrow [1,k]_\Z$, the $k$-track QTM
 $M'=(Q,\Si_{\pi(1)}\times\cdots\times\Si_{\pi(k)},\de')$
 is called the QTM constructed
 by the {\it permutation} $\pi$ of the tracks 
 (of $M$) if for any $(p,(\si_{\pi(1)},\ldots,\si_{\pi(k)}),q,
 (\ta_{\pi(1)},\ldots,\ta_{\pi(k)}),d)
 \in (Q\times(\Si_{\pi(1)}\times\cdots\times\Si_{\pi(k)}))^2
\times[-1,1]_\Z$, we have 
$$ 
\de'(p,(\si_{\pi(1)},\ldots,\si_{\pi(k)}),q,
(\ta_{\pi(1)},\ldots,\ta_{\pi(k)}),d)
=\de(p,(\si_1,\ldots,\si_k),q,(\ta_1,\ldots,\ta_k),d).
$$
 
\begin{Lemma}[Dovetailing lemma]\label{th:34}
 For $i=1,2$, let $M_i=(Q_i,\Si,\de_i)$ be an SNQTM
 with initial and final processor configurations $q_{i,0}$ and $q_{i,f}$.
 Then there is a normal form QTM $M=(Q,\Si,\de)$
 with initial and final processor configurations $q_{1,0}$ and $q_{2,f}$
 satisfying the following condition: 
 If $C_0$ is an initial configuration of $M_1$,
 the computation time for the input state $\ket{C_0}$ of $M_1$ is $s$, 
 and $M_{\de_1}^s\ket{C_0}=\sum_{T\in\Si^{\#}}\al_T\ket{q_{1,f},T,0}$, 
 then we have
\begin{eqnarray*} 
 M_{\de}^t\ket{C_0} &=& M_{\de_1}^t\ket{C_0}\ \ \mb{for}\ \ t<s,\\
 M_{\de}^{s+t}\ket{C_0} &=& \sum_{T\in\Si^{\#}}\al_T M_{\de_2}^t
 \ket{q_{2,0},T,0}\ \ \mb{for}\ \ t\ge 0.
\end{eqnarray*}
\end{Lemma}
Such an $M$ is called the QTM constructed by dovetailing $M_1$ and $M_2$.

Even if $M$ is the normal form QTM constructed by dovetailing
 SNQTMs $M_1$ and $M_2$, it is not always stationary. 
 What conditions ensure that the QTM $M$ is stationary?
 It is easy to see that one of the answers is
 to satisfy the following conditions (i) and (ii).

(i) The output state of $M_1$ for input state $\ket{q_0,{\rm tape}[x],0}$
 is represented by 
$$ 
\sum_{y\in\Si^n}\al_y\ket{q_{f},{\rm tape}[y],0} 
$$
 for some integer $n$, where $n$ depends on $|x|$.

(ii) $M_2$ is a stationary QTM such that
 if the input state is $\ket{q_0,{\rm tape}[x],0}$,
 the computation time for the input state depends only on $|x|$.

Condition (i) ensures that all computational basis vectors
 in the final superposition of $M_1$ represent
 the output strings of the same length, and
 condition (ii) ensures that
 if the final superposition of $M_1$ satisfying condition (i) is
 given as the initial state of $M_2$, every computational path 
 of $M_2$ reaches a final configuration simultaneously.
 These conditions are called the {\em dovetailing conditions}.

\begin{Lemma}[Branching lemma]\label{th:35} 
Let $M_i=(Q_i,\Si,\de_i)$ be an SNQTM for $i=1,2$. 
 Then there is an SNQTM $M=(Q,\Si\times\{B,1\},\de)$ satisfying
 the following condition with initial and final processor configurations
 $q_0$ and $q_f$.
 If the initial configuration of $M_i$ is $C_i=(q_{i,0},T_0,0)$
 such that the computation time of $M_i$ for $\ket{C_i}$ is $s_i$ and
 that $M_{\de_i}^{s_i}\ket{C_i}=\sum_{T\in\Si^{\#}}\al_{i,T}\ket{q_{i,f},T,0}$,
 then we have
$$
 M_\de^{s_i+4}|q_0,(T_0,T_i),0\r=
 \sum_{T\in\Si^{\#}}\al_{i,T}\ket{q_f,(T,T_i),0},
$$
 where $T_1={\rm tape}[B]$ and $T_2={\rm tape}[1]$. 
\end{Lemma}

\begin{Lemma}[Looping lemma]\label{th:36} 
There are an SNRTM $M=(Q,\Si,\de)$ and a constant $c$
 with the following properties. 
 On any positive input $k$ written in binary, 
 the computation time of $M$ is $t=O(k\log^ck)$ and 
 the output state of $M$ for the input state $\ket{q_0,T,0}$
 is $\ket{q_f,T,0}$. 
 Moreover, $M$ on input $k$ visits a special processor configuration $q^*$ 
 exactly $k$ times, each time with its head back in cell $0$.
 That is, there exist some $q^*$ in $Q$ and $k$ 
 positive integers $t_i<t$, where $i=1,\ldots,k$, such that 
$$ 
||E^{\hat{q}}(q^*)E^{\hat{\xi}}(0)M_\de^{t_i}\ket{q_0,T,0}||^2=1\ \ and\ \ 
||E^{\hat{q}}(q^*)M_\de^{s}\ket{q_0,T,0}||^2=0\ \ (s\neq t_1,\ldots,t_k).
$$
\end{Lemma}
 An RTM $M$ satisfying the above condition is called a looping machine.

 For any real number $\ep>0$, we denote by
 ${\rm Acc}(\ep)$ the least number $m$ satisfying $\frac{1}{2^m}\le\ep$.
 For convenience, we define ${\rm Acc}(0)=B$.
 Let $\widetilde{\C}=\{a+ib|\ a,b\in\Q\}$. 
 The {\em code} of an $m\times n$ matrix $M=(m_{ij})$
 with the components in $\widetilde{\C}$ is defined to be
 the list of finite sequences of numbers
 $\l\l x_{11},y_{11}\r,\l x_{12},y_{12}\r,\cdots,
\l x_{mn},y_{mn}\r\r$, where $x_{ij}=\mb{Re}(m_{ij})$ and
 $y_{ij}=\mb{Im}(m_{ij})$. 

Let $\cH$ be the Hilbert space spanned by the orthonormal system 
 $\cB=\{\ket{1},\ldots,\ket{n}\}$ and $\cL(\cH)$ be the set of
 all linear transformations on $\cH$.
 Let $e$ be a function
 mapping any $(U,\ep)\in \cL(\cH)\times\R_{\ge 0}$ to
 the following finite string $e(U,\ep)$:
 if $U$ has the matrix $A=(a_{ij})$ with $a_{ij}=\l i|U|j\r$, 
 then $e(U,\ep)$ is the code of $A'=(a'_{ij})$, 
 where $A'$ is the element of the set
 $\cX=\{B=(b_{ij})\ |\ b_{ij}\in\widetilde{\C},\ ||A-B||\le\ep\}$
 chosen uniquely by appropriate means.
 We call $e(U,\ep)$ the {\em $\ep$-approximate code of $U$}.
 Let $M$ be a multi-track QTM such that
 the alphabet of each track contains 0 and 1. 
 For some $U$ in $\cL(\cH)$,
 we say that {\em given the $\ep'$-approximate code,
 a QTM $M$ carries out $U$ with accuracy $\ep$
 (in $t$ steps on the first track)},
 if there is a unitary transformation $U'$ 
 such that $||U'-U||\le\ep$ and for any $\ket{j}\in\cB$ we have
$$
 M_\de^t\ket{q_0,{\rm tape}[j,e(U,\ep'),{\rm Acc}(\ep)],0}
 =\sum_{i=1}^n\ket{q_f,{\rm tape}[i,e(U,\ep'),{\rm Acc}(\ep)],0}
 \l i|U'|j\r. 
$$
 In particular, if $\ep=\ep'=0$ in the above condition, 
 we merely say that {\em $M$ carries out $U$ (in $t$ steps)}.
 Analogously we say that $M$ carries out $U$
 with accuracy $\ep$ in $t$ steps on the {\em $i$-th track}
 under appropriate modification of the above definition.
 
 The following theorem is a restricted version of
 the unitary theorem found by Bernstein and Vazirani \cite{BV97},
 but it serves our purpose.
 
\begin{Theorem}[Unitary theorem]\label{th:37} 
 Let $\cH$ be the Hilbert space spanned by the orthonormal system 
 $\cB=\{\ket{1},\ldots,\ket{n}\}$. Then there is a two-way SNQTM $M$
 that for any unitary transformation $U$ on $\cH$,
 given the $\frac{\ep}{4(10\sqrt{n})^n}$-approximate code,
 carries out $U$ with accuracy $\ep$ in time polynomial in $\frac{1}{\ep}$
 and the length of the input on its first track.
\end{Theorem}

\section{Quantum Circuits}

An element of $\{0,1\}^m$ is called a {\it bit string of length $m$} 
 or an {\em $m$-bit string}.
 For any $m$-bit string $x=x_1\cdots x_m$,
 the bit $x_i$ is called the {\em $i$-th bit} of $x$.  
 An $m$-input $n$-output {\it Boolean gate}
 is a function mapping $m$-bit strings to $n$-bit strings.
 An $n$-input $n$-output Boolean gate is called an $n$-bit Boolean gate.
 Suppose that $G$ is an $m$-input $n$-output Boolean gate. 
 An $n$-bit string $y_1\cdots y_n$ is called the {\it output} of $G$
 for {\em input} $x_1\cdots x_m$ if $G(x_1\cdots x_m)=y_1\cdots y_n$.
 A Boolean gate $G$ is said to be {\it reversible}
 if $G$ is a bijection. 
 For example, the Boolean gate $M_2(N)$ that for input $xy\in\{0,1\}^2$
 produces output $x(x+y\ \mb{mod}\ 2)\in\{0,1\}^2$ is a 2-bit
 reversible Boolean gate called the controlled not gate.
 The first bit is called the control bit,
 and the second bit is called the target bit.

\begin{center}
\unitlength 4mm
\begin{picture}(18,7)
\put(0,2){\line(1,0){1}}
\put(0,5){\line(1,0){1}}
\put(1,1){\framebox(4,5){$M_2(N)$}}
\put(5,2){\line(1,0){1}}
\put(5,5){\line(1,0){1}}
\put(7,3.5){$=$}
\put(9,5){\line(1,0){4}}
\put(9,2){\line(1,0){1}}
\put(12,2){\line(1,0){1}}
\put(11,3){\line(0,1){2}}
\put(10,1){\framebox(2,2){$N$}}
\put(11,5){\circle*{0.5}}
\end{picture}

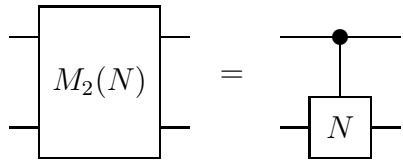
Figure 1 : The controlled not gate $M_2(N)$
\end{center}
    
To define quantum gates,
 we shall first introduce the notion of a wire. 
 A {\it wire} is an element of a countable set of 2-state systems.
 The set of wires is in one-to-one correspondence with 
 the set of natural numbers called {\it bit numbers}.
 Formally, the wire of bit number $j$ is represented
 by the Hilbert space $\cH_j\cong\C^2$
 spanned by a basis $\{\ket{0}_j,\ket{1}_j\}$,
 an orthonormal system in one-to-one correspondence with $\{0,1\}$.  
 An observable $\hat{n}_{j}=\ket{1}_j\bra{1}_j$
 in the Hilbert space $\cH_{j}$
 is called a $j$-th {\it bit observable}.
 Let $\La=\{j_{1},\ldots,j_{n}\}\subseteq \N$,
 where $j_{1}<\ldots<j_{n}$.
 A composite system of $n$ wires with different bit numbers in $\La$
 is represented by the Hilbert space $\cH_{\La}=\bigotimes_{j\in\La}\cH_{j}$.
 In the Hilbert space $\cH_\La$, the orthonormal system
$$
\{\ket{x_{1}}_{j_{1}}\cdots\ket{x_{n}}_{j_{n}}
|\ x_{1}\cdots x_{n}\in\{0,1\}^{n}\}
$$
 in one-to-one correspondence with $\{0,1\}^n$
 is called the {\it computational basis} on $\La$. 
Henceforth, we shall also write
 $\ket{x_{1},\ldots,x_{n}}=\ket{x_{1}}_{j_{1}}\cdots\ket{x_{n}}_{j_{n}}$.
 Thus, we obtain
$$
1\otimes\cdots 1\otimes{\hat n}_{j_{k}}\otimes1\cdots\otimes 1
\ket{x_{1},\ldots,x_k,\ldots,x_{n}}=x_{k}\ket{x_{1},\ldots,x_k,\ldots,x_{n}}.
$$

An $n$-bit quantum gate is physically
 to be interacting $n$ wires 
 such that the state transition
 from the input state to the output state
 is represented by the time evolution 
 of the composite system of the $n$ wires. 
 Formally, for any set $\La\subseteq\N$,
 a $\La$-{\it quantum gate} is defined to be 
 a unitary operator on the corresponding Hilbert space $\cH_{\La}$.
 In particular, a $[1,n]_\Z$-quantum gate is called
 an {\it $n$-bit quantum gate}. 
 The {\it S-matrix} of a $\La$-quantum gate 
 is the matrix representing its gate in the computational basis on $\La$. 
 For any $\La$-quantum gate $G$ and
 any unit vectors $\ket{\ps}$ and $\ket{\ph}$ in $\cH_{\La}$,
 if $G\ket{\ps}=\ket{\ph}$, 
 the vector $\ket{\ph}$ is called the {\it output state} of $G$
 for the {\em input state} $\ket{\ps}$. 
 In particular, if the input state is $\ket{\ps}=\ket{x_{1}\cdots x_{n}}$,
 the bit string $x_{1}\cdots x_{n}$ is called the {\it input} of $G$. 
 Henceforth when no confusion may arise, we usually identify 
 the S-matrix of a quantum gate with the quantum gate itself.

 We can represent an $n$-bit reversible Boolean gate
 by a $2^n\times 2^n$ orthogonal matrix
 whose entries are equal to zero or one.
 Thus we may consider an $n$-bit reversible Boolean gate 
 to be a sort of $n$-bit quantum gate, and
 consider that the class of reversible Boolean gates
 is a subclass of the class of quantum gates. 

Let $\pi$ be a permutation on $[1,n]_\Z$.
 The {\it permutation operator} of $\pi$
 is the operator $V_\pi$ on $\cH_{[1,n]_\Z}$
 that transforms $\ket{x_1\cdots x_n}$ to $|x_{\pi(1)},\ldots,x_{\pi(n)}\r$
 for any $n$-bit string $x_1\cdots x_n$. 
For any finite set $\La$, we denote by $I_{\La}$
 the identity operator on $\cH_{\La}=\otimes_{\la\in\La}\cH_{\la}$.
 For any $m$-bit quantum gate $G$, the {\em $n$-bit extension} of $G$
 is the $n$-bit quantum gate $G\otimes I_{[m+1,n]_\Z}$
 denoted by $G[n]$, where $m\le n$. 
For any set $\cG$ of quantum gates, 
 an $n$-bit quantum gate $G$ is said to be {\it decomposable
 by $\cG$}
 if there are $n_i$-bit quantum gates $G_i$ in $\cG$ with $n_i\le n$
 and permutations $\pi_i$ on $[1,n]_\Z$ satisfying 
\begin{equation}\label{eq:41}
 G=U_1\cdots U_m,\quad\mbox{where}\quad
 U_i=V_{\pi_i}^{\da}G_i[n]V_{\pi_i}
\end{equation}
for $i=1,2,\ldots,m$. In this case, 
$G$ is also said to be {\em decomposable by $m$ gates in} $\cG$.
The least number of such $m$ is called the {\em size of $G$ for} $\cG$.
 For any $\ep>0$, we say that $G$ is
 {\em decomposable by $\cG$ with accuracy} $\ep$,
 if $||G-U_1\cdots U_m||\le\ep$ is satisfied
 instead of Eq.\ (\ref{eq:41}).
 
A {\em universal set} is a set of quantum gates
 by which any quantum gate is decomposable with any accuracy. 
An {\em elementary gate} is an element of a given universal set. 
Henceforth, $R_{1,\th},R_{2,\th},$ and $R_{3,\th}$
 denote the 1-bit quantum gates whose S-matrices are given as follows.
$$ 
 R_{1,\theta}=\left( \begin{array}{cc} \cos\theta&-\sin\theta\\ \sin\theta&
\cos\theta \end{array}\right),\ \ R_{2,\theta}=\left( \begin{array}{cc}
e^{i\theta}&0\\ 0&1 \end{array}\right),\ \ R_{3,\theta}=
\left( \begin{array}{cc}
1&0\\ 0&e^{i\theta}
 \end{array}\right). 
$$
Barenco et al.\ \cite{BBC$^+$95} proved that
 any quantum gate is decomposable by the infinite set
$$ 
 \cG_u=\{R_{1,\th},R_{2,\th},R_{3,\th},M_2(N)|\ \th\in [0,2\pi]\}.
$$
as follows.
\begin{Theorem}\label{th:41} 
Any $n$-bit quantum gate $G$ is decomposable by 
 at most $O(n^32^{2n})$ quantum gates in $\cG_u$.
\end{Theorem}

Thus, $\cG_u$ is a universal set. 
 In what follows, the size of a quantum gate for $\cG_u$ 
 is merely called the {\it size} of the quantum gate.

We shall now consider a finite universal set. 
 Henceforth, $\cR$ denotes a polynomial time computable real
 $2\pi \sum_{i=1}^{\infty}2^{-2^i}$.
 The following lemma was obtained essentially
 by Bernstein and Vazirani \cite{BV97}.

\begin{Lemma}\label{th:42} 
For any $\th\in[0,2\pi]$ and $\ep>0$,
 there is a non-negative integer $k\le O(\frac{1}{\ep^4})$
 such that $|k\cR-\th|\ (${\rm mod}\ $2\pi)\le \ep$. 
 Moreover, there is a DTM
 which produces on input $\th\in\rP\R$ and ${\rm Acc}(\ep)$
 a non-negative integer $k\le O(\frac{1}{\ep^4})$
 satisfying the above inequality
 in time polynomial in the length of the input.
\end{Lemma}

Henceforth, $\cG_\cR$ denotes the finite set of quantum gates defined by
$$ 
 \cG_\cR=\{R_{1,\cR},R_{2,\cR},R_{3,\cR},
 M_2(N)|\ \cR=2\pi \sum_{i=1}^{\infty}2^{-2^i}\}. 
$$ 
Since any 1-bit quantum gate in $\cG_u$ is decomposable by $\cG_\cR$ 
 with any accuracy by Lemma \ref{th:42},
 the set $\cG_\cR$ is a universal set. 
In what follows, the size of a quantum gate for $\cG_\cR$
 is called the {\it $\cG_\cR$-size} of the quantum gate.

An $n$-bit quantum circuit consists of quantum gates and wires,
 and represents how those gates are connected
 with some of those wires. 
 Formally, it is defined as follows.
 Let $\cG$ be a set of quantum gates. 
 An $n$-bit {\it quantum circuit} $K$ {\it based on} $\cG$ 
 is a finite sequence $(G_m,\pi_m),\ldots,(G_1,\pi_1)$
 such that each pair $(G_i,\pi_i)$ satisfies the following conditions.

(1) $G_i$ is an $n_i$-bit quantum gate in $\cG$ with $n_i\le n$.

(2) $\pi_i$ is a permutation on $[1,n]_\Z$.\\
In this case, we say that the wire of bit number $\pi_i(j)$,
 where $j\le n_i$,
 {\em is connected with} the {\em $j$-th pin} of $G_i$. 
 The positive integer $m$ is called the {\em size of $K$ for} $\cG$. 
 In particular, the size of $K$ for $\cG_u$ is merely called
 the {\it size} of $K$
 and the size of $K$ for $\cG_\cR$ is called
 the {\it $\cG_\cR$-size} of $K$. 
 The unitary operator $U_m\cdots U_1$, 
 where $U_i=V_{\pi_i}^{\da}G_i[n]V_{\pi_i}$ for $i\in[1,m]_\Z$,
 is called the $n$-bit {\em quantum gate determined by} $K$
 and denoted by $G(K)$. From the definition, 
 the size of $G(K)$ for $\cG$ is at most the size of $K$ for $\cG$. 
 Suppose that $K_1=(G_{m},\pi_{m}),\ldots,(G_{1},\pi_{1})$
 and $K_2=(G'_{l},\pi'_{l}),\ldots,(G'_{1},\pi'_{1})$  
 are $n$-bit quantum circuits based on $\cG$. 
 Then $K_2\circ K_1=(G'_{l},\pi'_{l}),\ldots,(G'_{1},\pi'_{1}),
 (G_{m},\pi_{m}),\ldots,(G_{1},\pi_{1})$ is called 
 the concatenation of $K_1$ and $K_2$, and  
 $K_1^n=\underbrace{K_1\circ\cdots\circ K_1}_n$ is called 
 the concatenation of $n$ $K_1$'s.

Next, we define $k$-input $m$-output quantum circuits. 
 A $k$-input $m$-output $n$-bit quantum circuit is physically to be
 an $n$-bit quantum circuit based on a set of quantum gates;
 its input is a $k$-bit string and a constant $(n-k)$-bit string,
 and its output is the $m$-bit string
 obtained by measuring the bit observables of specified $m$ wires after 
 the unitary transformation determined by the circuit.
 
Formally, a $k$-{\it input} $m$-{\it output} $n$-bit {\it quantum circuit}
 $\bK$ is a 4-tuple $(K,\La_{1},\La_{2},S)$
 satisfying the following conditions.

(1) $K$ is an $n$-bit quantum circuit.

(2) $\La_{1}$ and $\La_{2}$ are two subsets of $[1,n]_\Z$
 satisfying $|\La_{1}|=k$ and $|\La_{2}|=m$, respectively.

(3) $S$ is a function from $[1,n]_\Z\setminus\La_{1}$ to $\{0,1\}$.\\ 
Henceforth, we write $b_j=S(j)$ 
for any $j\in [1,n]_\Z\setminus\Lambda_{1}$.
 
Let $\bK=(K,\La_{1},\La_{2},S)$ be a $k$-input $m$-output 
 $n$-bit quantum circuit, where $\La_{1}=\{j_{1},\ldots,j_{k}\}$ and
 $\La_{2}=\{i_{1},\ldots,i_{m}\}$, 
 and let $u=u_{1}\cdots u_{n}$ be the $n$-bit string satisfying
 $u_{j_1}=x_{1},\ldots,u_{j_k}=x_k$ 
 for a $k$-bit string $x=x_{1}\cdots x_{k}$ and 
 $u_{j}=b_{j}$ for all $j\in[1,n]_\Z\setminus\La_{1}$. 
In what follows, the $n$-bit string $u$ obtained by such construction
 is denoted by $u(x,\bK)$. 
Let $\ket{\ph}$ be the output state of $G(K)$ for input $u(x,\bK)$. 
 If the bit observables ${\hat n}_{i_{1}},\ldots,{\hat n}_{i_{m}}$
 are measured simultaneously in the output state $\ket{\ph}$,
 and the outcomes of these measurements are $y_1,\ldots,y_m$,
 then the bit string $y=y_{1}\cdots y_{m}$ is considered as
 the {\em output} of $\bK$ for {\em input} $x$. 
 From the statistical formula of quantum physics,
 the probability $\rh^K(y|x)$ such that
 $y$ is the output of $\bK$ for input $x$ is represented by 
$$ 
 \rh^K(y|x)=\l u(x,\bK)|{G(K)}^{\da}E_{i_1}(y_{1})
\cdots E_{i_m}(y_{m}){G(K)} |u(x,\bK)\r,  
$$
 where $E_{i_p}(y_p)$ is the spectral projection of 
 $1\otimes\cdots 1\otimes\hat{n}_{i_p}\otimes 1\cdots\otimes 1$ 
 pertaining to its eigenvalue $y_p$. 
 We can consider that $\bK$ associates each $k$-bit string $x$
 with the probability distribution $\rh^K(\cdot|x)$ on $\{0,1\}^m$. 
 The distribution $\rh^K(\cdot|x)$ is called
 the {\em output distribution for} $x$ {\em determined by} $\bK$. 
 Henceforth, when no confusion may arise,
 we shall identify $\bK$ with $K$.

Now, we shall give the notion of a simulation of a QTM
 by a quantum circuit.
 The total variation distance
 between two distributions $\cD$ and $\cD'$
 over the same domain $I$ is $\sum_{i\in I}|\cD(i)-\cD'(i)|$.
 A quantum circuit $K$ will be said to $t$-simulate a QTM $M=(Q,\Si,\de)$
 with accuracy $\ep$, if the following holds for any $\Si$-string $x$.
 Let $\cD$ be the probability distribution of the outcomes
 of the simultaneous measurement of the tape cells from cell
 $-t$ to cell $t$ after $t$ steps of $M$
 for input state $\ket{q_0,{\rm tape}[x],0}$. 
 Let $\cD'$ be the probability distribution of
 the $\Si$-string obtained by decoding the output of $K$ for the input
 of the bit string obtained by encoding $x$.
 Then the total variation distance between $\cD$ and $\cD'$ is at most $\ep$.
 Formally, it is defined as follows.

Let $e:\Si\rightarrow\{0,1\}^{\la}$,
 where $\la=\lceil \log |\Si| \rceil$,
 be an injection computable in polynomial time,
 and let $d:\{0,1\}^{\la}\rightarrow\Si$
 be a function computable in polynomial time
 such that $d\cdot e=\mbox{id}$.
 For any $\Si$-string $x=x_1\cdots x_k$, positive integer $t$ and
 bit string $z=z_1\cdots z_{2t+1}$, where $z_i\in\{0,1\}^{\la}$, 
 we define the encoding function
 $e_t:\Si^*\rightarrow\{0,1\}^{(2t+1)\la}$ by
$$
e_t(x_1\cdots x_k)=\left\{
\begin{array}{ll}
\underbrace{e(B)\cdots e(B)}_{t}e(x_1)\cdots e(x_k)
\underbrace{e(B)\cdots e(B)}_{t+1-k}&\mbox{if}\ t+1\ge k,\\
\underbrace{e(B)\cdots e(B)}_{t}e(x_1)\cdots e(x_{t+1})&\mbox{if}\ t+1<k,
\end{array}\right.
$$
 and define the decoding function
 $d_t:\{0,1\}^{(2t+1)\la}\rightarrow\Si^{2t+1}$ by
$$
 d_t(z_1\cdots z_{2t+1})= d(z_1)\cdots d(z_{2t+1}). 
$$
 Then a ($(2t+1)\la$-input $(2t+1)\la$-output) quantum circuit $K$
 is said to $t$-{\it simulate} a QTM $M=(Q,\Si,\de)$
 with {\it accuracy} $\ep$
 (under the {\it encoding} $e_t$ and the {\it decoding} $d_t$), 
 if for any $\Si$-string $x$, we have
$$ 
\sum_{y\in\Si^{2t+1}}\left|\rh_{t}^M(y|x)-
 \tilde{\rh}^K(y|x)\right|\le\ep, 
$$ 
where 
\begin{eqnarray*}
 \tilde{\rh}^K(y|x) &=& \sum_{z\in d_{t}^{-1}(y)} \rh^K(z|e_t(x)),\\
 \rh_{t}^M(y|x) &=&
 \l q_0,{\rm tape}[x],0|(M_\de^t)^{\da}E_{M,-t}(y_{1})\cdots E_{M,t}(y_{2t+1})
 M_\de^t|q_0,{\rm tape}[x],0\r.
\end{eqnarray*}
When $\ep=0$, the quantum circuit $K$ is merely said to
 $t$-{\it simulate} the QTM $M$.

 Yao \cite{Yao93} discussed
 the simulation of a QTM by a quantum circuit
 under a similar but different formulation.
 He showed that given a QTM $M=(Q,\Si,\de)$
 and positive integers $t$ and $n$,
 there is an $n$-input quantum circuit that simulates $M$
 for $t$ steps on any input of $M$ 
 with length $\lceil n/\lceil\log |\Si|\rceil\rceil$ and 
 that its ``size''(the ``size'' is the number of Deutsch gates
 \cite{Deu89} constructing the circuit)
 is at most some fixed polynomial in $t$ and $n$. 
 Our formulation requires that
 a quantum circuit simulate a QTM $M$
 on {\it every} input of $M$
 and we shall extend quantum circuits used
 by Yao \cite{Yao93} to those which can simulate multi-tape QTMs. 
 In addition, we shall construct
 a quantum circuit based on $\cG_u$ instead of Deutsch gates
 in order to take advantage of this simulation later.

\begin{Theorem}\label{th:43}
 Let $M=(Q,\Si,\de)$ be a $k$-tape QTM, and let $t\in\N$. 
 Then, there is a quantum circuit of size $O(t^{k+1})$
 that $t$-simulates $M$.
\end{Theorem}

{\it Proof.} 
 We consider the case where $M$ is a single tape QTM. 
 See appendix A for the generalization to multi-tape QTMs.
 We shall construct a quantum circuit $K_\cG$
 which $t$-simulates $M$.
 The quantum gate determined by $K_\cG$
 is connected with $l_0+(2t+1)l$ wires,
 where $l_0=\lceil \log |Q|\rceil$ and
 $l=2+\lceil \log |\Si| \rceil$.
 We divide their wires into a part consisting of
 the first $l_0$ wires and $2t+1$ parts
 which are respectively consisting of $l$ wires.
 The part consisting of the first $l_0$ wires
 represents the processor configuration of $M$.
 This set of wires is called cell `P' of $K_\cG$.
 The state of cell P of $K_\cG$ is represented by
 a unit vector in the Hilbert space 
 spanned by the computational basis $\{\ket{q}\}$,
 where $q\in\{0,1\}^{l_0}$. 
 For $j\in[0,2t]_\Z$,
 the wires of bit numbers $l_0+jl+1,\ldots,l_0+jl+l$
 represent the symbol in the $(j-t)$-th cell of $M$ and
 whether the head scans this cell or not.
 This set of wires is called cell $j-t$ of $K_\cG$.
 For $i\in[-t,t]_\Z$, the state of cell $i$ of $K_\cG$
 is represented by a unit vector in the Hilbert space 
 spanned by the computational basis $\{\ket{\si_is_i}\}$,
 where $\si_i\in\{0,1\}^{\lceil \log |\Si| \rceil}$
 and $s_i\in\{0,1\}^{2}$. 

Next, we define quantum gates $G_1$ and $G_2$,
 two types of components of $K_\cG$. 
 In what follows, $p,q,\ldots$ denote binary strings
 representing elements of $Q$, the symbols
 $\si,\ta,\ldots$ denote binary strings
 representing elements of $\Si$,
 and $s=\bar{0},\bar{1},\bar{2}$ denote 00,01,10, respectively. 
 Then we denote the computational basis state
 $\ket{q\si_1s_1\si_2s_2\cdots \si_ks_k}$
 on the set $[1,l_0+kl]_{\Z}$ of bit numbers
 by $\ket{q;\si_1s_1;\si_2s_2;\cdots;\si_ks_k}$.
 Now $G_1$ is an $(l_0+3l)$-bit quantum gate satisfying
 the following conditions (i) and (ii).

(i) $G_1\ket{w_{p,\si_1,\si,\si_3}}
=\ket{v_{p,\si_1,\si,\si_3}}$, where 
\begin{eqnarray*}
 \ket{w_{p,\si_1,\si,\si_3}}
 &=&\ket{p; \si_1\bar{0};\si \bar{1}; \si_3\bar{0}},\\
 \ket{v_{p,\si_1,\si,\si_3}}
 &=&\sum_{q,\ta}\de(p,\si,q,\ta,-1)
    \ket{q; \si_1\bar{2};\ta \bar{0};\si_3\bar{0}}
   +\sum_{q,\ta}\de(p,\si,q,\ta,0)
    \ket{q; \si_1\bar{0};\ta \bar{2};\si_3\bar{0}}\\
 & &\mbox{ }
   +\sum_{q,\ta}\de(p,\si,q,\ta,1)
    \ket{q; \si_1\bar{0};\ta \bar{0};\si_3\bar{2}}
\end{eqnarray*}
 for any $(p,\si_1,\si,\si_3)\in Q\times\Si^3$; 
 the summation $\sum_{q,\ta}$ is taken
 over all $(q,\ta)\in Q\times \Si$.

(ii) $G_1\ket{h}=\ket{h}$ for each vector $\ket{h}$
 in the subspace $H$ of $\C^{2^{l_0+3l}}$
 spanned by three types of vectors:

(1) $\ket{q;\si_1s_1;\si_2s_2;\si_3s_3}$,\\
where $s_2\neq\bar{1}$ and none of $s_1,s_2,s_3$
 are equal to $\bar{2}$;

\medskip
(2) $\ket{u^1_{p,\si,\si_2,\si_3}}
=\sum_{q,\ta}\de(p,\si,q,\ta,0)
\ket{q; \ta\bar{2};\si_2\bar{0};\si_3\bar{0}}
+\sum_{q,\ta}\de(p,\si,q,\ta,1)
\ket{q; \ta\bar{0};\si_2\bar{2};\si_3\bar{0}}$;

\medskip
(3) $\ket{u^2_{p,\si,\ta,\si_1,\si_2,\si_3}}
=\sum_{q\in Q}\de(p,\si,q,\ta,1)
\ket{q; \si_1\bar{2};\si_2\bar{0};\si_3\bar{0}}$.

\medskip
Let $W=\{\ket{w_{p,\si,\si_1,\si_3}}|\ (p,\si,\si_1,\si_3)\in
 Q\times\Si^3\}^{\bot\bot}$ and 
 $V=\{\ket{v_{p,\si,\si_1,\si_3}}|\ (p,\si,\si_1,\si_3)\in
 Q\times\Si^3\}^{\bot\bot}$, 
 where $S^{\bot}$ denotes the orthogonal complement of a set $S$ so that 
 $S^{\bot\bot}$ denotes the subspace generated by $S$. 
By Theorem \ref{th:21} the subspaces $W$, $V$ and $H$
 are all orthogonal one another
 and it is verified that
 $\{\ket{v_{p,\si,\si_1,\si_3}}\}$
 is an orthonormal system of $V$.
 Thus, there exists a quantum gate $G_1$
 satisfying the above condition. 
Let $G_2$ be an $(l_0+(2t+1)l)$-bit reversible Boolean gate
 which changes all $s_i=\bar{2}$ to $s_i=\bar{1}$.

Henceforth, given any $m\in[1,2t+1]_\Z$, 
 we say that an $(l_0+ml)$-bit quantum gate $G$
 is connected with cells $i_1,\ldots,i_m$,
 where $i_1<\cdots <i_m$, 
 if each $j_0$-th pin of $G$, for $j_0\in[1,l_0]_\Z$,
 and each $(l_0+jl-l+k)$-th pin of $G$, for $j\in[1,m]_\Z,\ k\in[1,l]_\Z$,
 are respectively connected with
 the wires of bit numbers $j_0$ and $l_0+(i_j+t)l+k$. 
 Now let $K_\cG$ be the quantum circuit based on 
 $\cG=\{G_1,G_2\}$ constructed as follows. 
 First, $2t-1$ $G_1$'s are connected in such a way that
 for $j\in[1,2t-1]_\Z$
 the $j$-th $G_1$ is connected with
 cells $j-t-1,j-t$ and $j-t+1$. 
 The $(l_0+(2t+1)l)$-bit quantum circuit
 constructed from these $G_1$'s is called $K_1$. 
 Lastly, $G_2$ is connected with cells $-t,-t+1,\ldots,t$. 
 The $(l_0+(2t+1)l)$-bit quantum circuit
 constructed from this $G_2$ is called $K_2$.
 Let $K_\cG=(K_2\circ K_1)^t$. 
 The quantum circuit $K_2\circ K_1$
 is illustrated in Figure 2. 
 From the definitions of $G_1$ and $G_2$,
 it can be verified that $K_\cG$ carries out
 the operation corresponding to one step of $M$ as follows.

\begin{center}\unitlength 3mm
\begin{picture}(47,22)
\put(8,3){\line(0,1){2}}\put(9,3){\line(0,1){2}}
\put(12,3){\line(0,1){2}}\put(13,3){\line(0,1){2}}
\put(16,3){\line(0,1){2}}\put(17,3){\line(0,1){2}}
\put(20,3){\line(0,1){4}}\put(21,3){\line(0,1){4}}
\put(41,3){\line(0,1){12}}\put(42,3){\line(0,1){12}}
\put(7,5){\framebox(11,1)}

\put(8,6){\line(0,1){11}}\put(9,6){\line(0,1){11}}
\put(12,6){\line(0,1){1}}\put(13,6){\line(0,1){1}}
\put(16,6){\line(0,1){1}}\put(17,6){\line(0,1){1}}
\put(11,7){\framebox(11,1)}
\put(12,8){\line(0,1){9}}\put(13,8){\line(0,1){9}}
\put(16,8){\line(0,1){1}}\put(17,8){\line(0,1){1}}
\put(20,8){\line(0,1){1}}\put(21,8){\line(0,1){1}}
\put(33,16){\line(0,1){1}}\put(34,16){\line(0,1){1}}
\put(33,14){\line(0,1){1}}\put(34,14){\line(0,1){1}}
\put(33,3){\line(0,1){7}}\put(34,3){\line(0,1){7}}
\put(37,3){\line(0,1){10}}\put(38,3){\line(0,1){10}}
\put(28,13){\framebox(11,1)}
\put(32,15){\framebox(11,1)}
\put(41,16){\line(0,1){1}}\put(42,16){\line(0,1){1}}
\put(37,14){\line(0,1){1}}\put(38,14){\line(0,1){1}}
\multiput(23,5)(2,0){5}{\line(1,0){0.1}}
\multiput(17,11)(2,0){10}{\line(1,0){0.1}}
\multiput(17,19)(2,0){7}{\line(1,0){0.1}}
\put(37,16){\line(0,1){1}}\put(38,16){\line(0,1){1}}
\put(2,17){\framebox(41,1)}
\put(8,18){\line(0,1){1}}\put(9,18){\line(0,1){1}}
\put(41,18){\line(0,1){1}}\put(42,18){\line(0,1){1}}
\put(0,1){cell}\put(3.5,1){P}
\put(7.5,1){$-t$}\put(10,1){$-t+1$}\put(36.5,1){$t-1$}\put(41.5,1){$t$}
\put(0,5){$G_1$}\put(0,7){$G_1$}\put(0,17){$G_2$}

\put(2,5){\framebox(3,1)}\put(2,7){\framebox(3,1)}
\put(2,13){\framebox(3,1)}\put(2,15){\framebox(3,1)}
\put(5,5.5){\line(1,0){2}}\put(5,7.5){\line(1,0){6}}
\put(5,13.5){\line(1,0){23}}\put(5,15.5){\line(1,0){27}}

\put(3,3){\line(0,1){2}}\put(4,3){\line(0,1){2}}
\put(3,6){\line(0,1){1}}\put(4,6){\line(0,1){1}}
\put(3,8){\line(0,1){1}}\put(4,8){\line(0,1){1}}
\put(3,14){\line(0,1){1}}\put(4,14){\line(0,1){1}}
\put(3,16){\line(0,1){1}}\put(4,16){\line(0,1){1}}
\put(3,18){\line(0,1){1}}\put(4,18){\line(0,1){1}}
\end{picture}

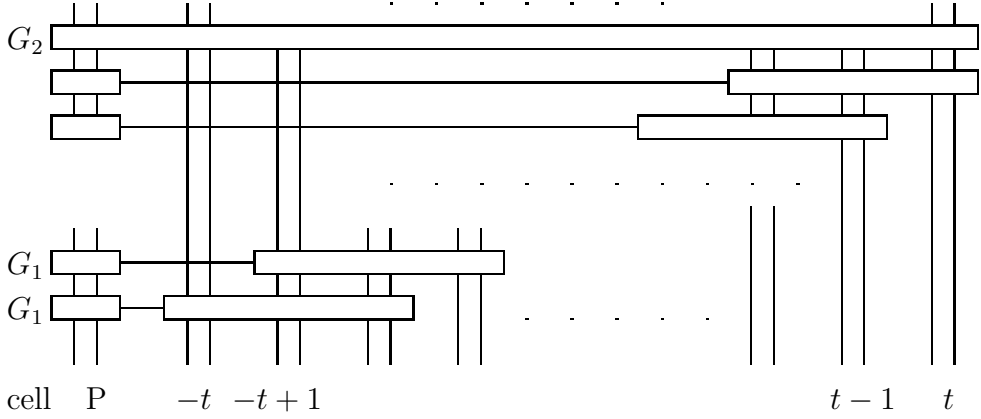
Figure 2 : The quantum circuit $K_2\circ K_1$ based on $\cG$
\end{center}

If the state of $M$ after $t'$ steps with $t'<t$
 is $\ket{p,T,i}$ with $T(i)=\si$,
 the input state of the $(t'+1)$-th $K_2\circ K_1$ is 
$$ 
\ket{p; T(-t)\bar{0};\cdots; T(i-1)\bar{0};T(i)\bar{1};
 T(i+1)\bar{0};\cdots; T(t)\bar{0}}. 
$$
From condition (ii-1) of $G_1$,
 this state does not change
 until $i$-th $G_1$ is carried out.
 When $i$-th $G_1$ is carried out,
 from condition (i) of $G_1$
 this state is transformed into
$$ 
\begin{array}{l}
 \sum_{q,\ta}\de(p,\si,q,\ta,-1)
 \ket{q; T(-t)\bar{0};\cdots ; T(i-1)\bar{2};
 \ta\bar{0}; T(i+1)\bar{0}; \cdots; T(t)\bar{0}}\\
 +\sum_{q,\ta}\de(p,\si,q,\ta,0)
 \ket{q; T(-t)\bar{0};\cdots ; T(i-1)\bar{0};
 \ta\bar{2}; T(i+1)\bar{0}; \cdots; T(t)\bar{0}}\\
 +\sum_{q,\ta}\de(p,\si,q,\ta,1)
 \ket{q; T(-t)\bar{0};\cdots ; T(i-1)\bar{0};
 \ta\bar{0};T(i+1)\bar{2}; \cdots; T(t)\bar{0}}. 
\end{array} 
$$  
By condition (ii) of $G_1$
 this state does not change until $K_2$ is carried out.
 Finally, from the definition of $G_2$, 
 the state after passing $K_2$
 in the $(t'+1)$-th $K_2\circ K_1$ is transformed into 
$$ 
\begin{array}{l}
 \sum_{q,\ta}\de(p,\si,q,\ta,-1)
 \ket{q; T(-t)\bar{0};\cdots ;T(i-1)\bar{1};\ta \bar{0};
 T(i+1)\bar{0}; \cdots; T(t)\bar{0}}\\
 +\sum_{q,\ta}\de(p,\si,q,\ta,0)
 \ket{q; T(-t)\bar{0};\cdots ;T(i-1)\bar{0};\ta \bar{1};
 T(i+1)\bar{0}; \cdots; T(t)\bar{0}}\\
 +\sum_{q,\ta}\de(p,\si,q,\ta,1)
 \ket{q; T(-t)\bar{0};\cdots ;T(i-1)\bar{0};\ta \bar{0};
 T(i+1)\bar{1}; \cdots; T(t)\bar{0}}.
\end{array} 
$$ 
 By the above transformation, it can be verified that
 $K_2\circ K_1$ simulates the operation of $M$ such that 
 ``if the processor configuration is $p$ 
 and the head reads the symbol $\si$ of cell $i$ after $t'$ steps, 
 then the head writes the symbol $\ta$, 
 the processor configuration turns to $q$, 
 and the head moves to $d$ with amplitude $\de(p,\si,q,\ta,d)$''.

From Theorem \ref{th:41} the quantum gate $G_1$
 is decomposable by $O(1)$ gates in $\cG_u$.
 It is easy to see that
 the quantum gate $G_2$ is decomposable by
 $O(2t+1)$ gates in $\cG_u$.
 Thus there are an $(l_0+3l)$-bit quantum circuit $K_{u,1}$
 of constant size
 and an $(l_0+(2t+1)l)$-bit quantum circuit $K_{u,2}$
 of size $O(2t+1)$ based on $\cG_u$
 such that the quantum gates determined by them
 are $G_1$ and $G_{2}$, respectively. 
 Now let $K_a$ be an $(l_0+(2t+1)l)$-bit quantum circuit
 obtained by decomposing each $G_1$ in $K_1$
 into $O(1)$ gates in $\cG_u$.
 Then the size of $K_a$ is $O(2t+1)$.
 Similarly, from $K_2$
 we can obtain an $(l_0+(2t+1)l)$-bit quantum circuit
 $K_b$ of size $O(2t+1)$. Thus, $K=(K_b\circ K_a)^t$
 is a quantum circuit of size $O(t^2)$
 that $t$-simulates $M$.\ \ \ \qed

Let $V$ be a $2^n$-dimensional transformation
 and $A=\{j_1,\ldots,j_n\}$
 a set of integers with $j_1<\cdots<j_n$. 
Then we say that a multi-track QTM $M$ {\em carries out
 $V$ (with accuracy $\ep$) on the cell-set} $A$
 of the $i$-th track,
 if $M$ carries out the following algorithm. 
 
1. For $m=1,\ldots,n$, the QTM $M$ transfers
 the symbol written on each cell $j_m$ of the $i$-th track
 to cell $m$ of an empty extra track.
 Henceforth, let this extra track be the $k$-th track.

2. $M$ carries out $V$ (with accuracy $\ep$)
 on the $k$-th track.

3. $M$ transfers
 the symbol written on each cell $m$ of the $k$-th track
 to cell $j_m$ of the $i$-th track.

 Now, we give a proof of the existence of a universal QTM 
 that simulates every QTM in $\rP\C$ efficiently
 with arbitrary given accuracy. 

\begin{Theorem}\label{th:44} 
 There is a two-way SNQTM $M_u$ such that
 for any positive integer $t$, positive number $\ep$,
 QTM $M$ in $\rP\C$, and input string $x$ of $M$,
 the QTM $M_u$ on input $(t,\ep,c_M,x)$
 simulates $M$ for $t$ steps with accuracy $\ep$
 and slowdown of at most a polynomial in $t$ and $\frac{1}{\ep}$,
 where $c_M$ is the code of $M$.
\end{Theorem}

{\it Proof.}\ \ For simplicity, we consider the case where
 $M$ is a single tape QTM. When $M$ is a multi-tape QTM,
 we can prove this theorem similar to the proof shown
 in the following by using a quantum circuit given in appendix A
 instead of a quantum circuit given in the proof of Theorem \ref{th:43}. 

 In what follows, we shall construct a multi-track QTM
 $M_u=(Q_u,\Si_u,\de_u)$ that simulates $M$ for $t$ steps
 with accuracy $\ep$ for any given $t$, $\ep$, and $M$.
 The input of $M_u$ consists of
 the input $x$ of $M$, the desired number of steps $t$,
 the desired accuracy $\ep$, and the code of $M$.
 Henceforth, we fix $t$, $\ep$, and $M$. In this proof,
 we shall use the same notations as in the proof
 of Theorem \ref{th:43}. By the proof of Theorem \ref{th:43}, 
 there is a quantum circuit $K_\cG=(K_2\circ K_1)^{t}$ based on
 $\cG=\{G_1,G_2\}$ that $t$-simulates $M$.
 The QTM $M_u$ has six tracks and
 the alphabet of each track contains 0 and 1.
 The first track of $M_u$ will be used
 to represent the computation of $K_\cG$ approximately.
 The second and the third track of $M_u$
 will be respectively used to record an approximate code of $G_1$
 and ${\rm Acc}(\ep)$.
 The fourth track of $M_u$ will contain
 counters $C_0$ and $C_1$.
 The values of $C_0$ and $C_1$ count
 the numbers of subcircuits of the form $K_2\circ K_1$ and
 $G_1$ in $K_\cG$ which have been carried out so far, respectively. 
 The fifth track of $M_u$ is used to record the input of $M_u$.
 The sixth track is used as a working track.  

 Let $k=2^{l_0+3l}$ and $\ep'\le\frac{\ep}{16t(2t-1)(10\sqrt{k})^k}$.
 The QTM $M_u$ carries out $K_\cG$ with accuracy $\ep$
 after a preparation.
 The preparation is to compute the $\ep'$-approximate code $c(G_1)$
 of $G_1$ from $c_M$ and write $c(G_1)$ on the second track of $M_u$,
 to write ${\rm Acc}(\ep)$ on the third track of $M_u$,
 and to write the $(l_0+(2t+1)l)$-bit string 
 $x'=q_0 T(-t)\bar{0}\cdots T(0)\bar{1}\cdots T(t)\bar{0}$
 corresponding to the initial configuration
 $\ket{q_0,{\rm tape}[x],0}$ of $M$ on the first track of $M_u$, 
 where the string $x'$ represents the input of $K_\cG$.
 Given $c_M$ and a sufficiently small positive number $c\ep'$,
 where $c$ depends only on $k$, i.e., $|Q|$ and $|\Si|$,
 but is independent of $t$ and $\ep$,
 we can compute the $\ep'$-approximate code of
 the ($k$ dimensional) S-matrix of $G_1$ in polynomial time
 in $\log t$ and $\log\frac{1}{\ep}$
 by using the definition of $G_1$ given in the proof of Theorem \ref{th:43}
 and the orthonormalization of Schmidt.
 By the synchronization theorem
 there is an SNQTM that carries out the preparation given above
 in time polynomial in the length of the input of $M_u$, i.e.,
 a polynomial in $t$ and $\log\frac{1}{\ep}$.

The algorithm for carrying out $K_\cG$ with accuracy $\ep$ is as follows. 
 At first, the values of counters $C_0$ and $C_1$ are zero.

Step 1. Carry out steps 2--4
 until the value of counter $C_0$ comes to $t$. 

Step 2. Carry out steps 2.1 and 2.2
 until the value of counter $C_1$ comes to a multiple of $2t-1$.

Step 2.1. When the value of $C_1$ is $i$ (mod $2t-1$),
 carry out the $2^{l_0+3l}$-dimensional transformation $G_1$
 with accuracy $\frac{\ep}{4t(2t-1)}$
 on the cell-set $[0,l_0-1]_\Z\cup[l_0+il,l_0+il+l-1]_\Z$
 of the first track.

Step 2.2. Increase the value of counter $C_1$ by one.

Step 3. Carry out the $2^{l_0+(2t+1)l}$-dimensional transformation
 $G_2$ on the cell-set $[0,l_0+(2t+1)l-1]_\Z$ of the first track.

Step 4. Increase the value of counter $C_0$ by one.
  
Since $G_2$ is a reversible Boolean gate
 which transforms all $s_i=\bar{2}$ to $s_i=\bar{1}$,
 we can construct an SNRTM $M_2$ that carries out step 3
 in time polynomial in $t$ by the synchronization theorem.
 We can construct an SNQTM $M_1$ that carries out
 $G_1$ with accuracy $\frac{\ep}{4t(2t-1)}$
 in time polynomial in $\frac{4t(2t-1)}{\ep}$ and
 $|c(G_1)|=O(k^2\log \frac{1}{\ep'})$ by the unitary theorem.
 Moreover, we can construct SNQTMs to run
 counters $C_0$ and $C_1$ by the looping lemma.
 The QTM $M_u$ can be constructed by applying
 the addition of tracks, the permutation of tracks,
 and the dovetailing lemma to the above SNQTMs and
 the SNQTM that carries out the preparation.

It is clearly verified that the operation of steps 2--4 
 corresponds to carrying out $K_2\circ K_1$ with
 accuracy $\frac{\ep}{4t}$ by the proof of Theorem \ref{th:43}. 
 Thus if the value of counter $C_0$ comes to $t$, then
 $M_u$ carries out the quantum gate determined by
 $K_\cG$ with accuracy $\ep$
 (It is known that if $||\ket{\phi}-\ket{\psi}||\le\ep$
 for two state vectors $\ket{\phi},\ket{\psi}$,
 the total variation distance between the probability distributions
 determined by them is at most $4\ep$ \cite{BV97}).
 Now let $q_{0,u}$ and $q_{f,u}$ be the initial and
 final processor configurations of $M_u$.
 Let the encoding
 $e:\cC(Q,\Si)\rightarrow\cC(Q_u,\Si_u)$ be a function
 satisfying $e(q_0,{\rm tape}[x],0)=
(q_{0,u},{\rm tape}[B,B,B,B,\l t,\ep,c_M,x\r],0)$.
 Let $x=x_0x_1\cdots x_{|x|-1}$.
 Let the decoding $d:\cC(Q_u,\Si_u)\rightarrow\cC(Q,\Si)$ be a function
 satisfying the following condition.
 For any $(T^1,\ldots,T^6)\in\Si_u^\#$
 satisfying 
\begin{eqnarray*}
 & &T^1={\rm tape}[q T(-t)\bar{0}\cdots T(\xi)\bar{1}\cdots T(t)\bar{0}],\\
 & &T^2={\rm tape}[c(G_1)],
\ \ T^3={\rm tape}[{\rm Acc}(\ep)],
\ \ T^5={\rm tape}[\l t,\ep,c_M,x\r],
\end{eqnarray*} 
 the equation $d(q_{f,u},(T^1,\ldots,T^6),0)=(q,T',\xi)$ holds;
 the tape configuration $T'$ of $M$ satisfies 
$$ 
T'(i)=\left\{
\begin{array}{ll}
T(i) &\mbox{if}\ i\in[-t,t]_\Z,\\
x_i  &\mbox{if}\ t<|x|-1\ \mb{and}\ i\in[t+1,|x|-1]_\Z,\\
B    &\mb{otherwise}.
\end{array}\right.
$$
 It is easy to see that $M_u$ simulates $M$ for $t$ steps
 with accuracy $\ep$ and the computation time is bounded by
 a polynomial in $t$ and $\frac{1}{\ep}$.\ \ \ \qed 

{\em Remark 1.}\ Any pair of QTMs dovetailed
 in the proof of Theorem \ref{th:44} can be constructed so that
 it can satisfy the dovetailing conditions (cf.\ Lemma \ref{th:33}). 
 Thus, stationarity is preserved by dovetailing them. 
 Indeed, the fact that all QTMs constructed
 in the proof of Theorem \ref{th:44}
 satisfy conditions (i) and (ii) of the dovetailing conditions
 can be verified from the statements of the synchronization theorem,
 the looping lemma, and the unitary theorem.

{\em Remark 2.}\ Using a universal set different from us,
 Kitaev \cite{Kit97} and Solovay \cite{Sol99} independently proved
 lately that there is a quantum algorithm
 which decomposes a given $n$-bit quantum gate
 into poly$(2^n,\log\frac{1}{\ep})$ elementary gates
 with accuracy $\ep$. Applying this result to
 the proof of Theorem \ref{th:44},
 we can replace a polynomial in $n$ and $\frac{1}{\ep}$
 in the statement of Theorem \ref{th:44}
 by a polynomial in $n$ and $\log\frac{1}{\ep}$.  

\section{Computational complexity of uniform QCFs and QTMs}

A {\it quantum circuit family (QCF)} is an infinite sequence
 $\cK=\{\bK_n\}_{n\ge 1}$ such that $\bK_n$ is an $n$-input 
 ($f(n)$-output $g(n)$-bit) quantum circuit.
 A QCF $\cK$ is said to be {\em based on} a set $\cG$ of quantum gates
 if every $\bK_n$ in $\cK$ is based on $\cG$.
 A QCF $\cK$ is said to be of {\it size} $s$ based on $\cG$
 if the size of $\bK_n$ for $\cG$ is $s(n)$
 for a function $s$ from $\N$ to $\N$. 
 If $s$ is a polynomial, it is called
 a {\it polynomial size} QCF {\it based on} $\cG$.
 Moreover, if $\cG=\cG_u$,
 then $\cK$ is merely called a {\it polynomial size} QCF.  
 For any quantum circuit $K$, the quantum gate $G(K)$
 determined by $K$ is decomposable by $\cG_u$ from Theorem \ref{th:41}. 
 Thus, in what follows, we consider only quantum circuits based on
 subsets of $\cG_u$.

First we define the code of a quantum circuit based on $\cG_\cR$.
 Let $K=(G_m,\pi_m),\ldots,(G_1,\pi_1)$ be
 a quantum circuit based on $\cG_\cR$.
 Then the {\it $\cG_\cR$-code} of $K$, denoted by $c_r(K)$,
 is defined to be the list of finite sequences of natural numbers
 $\l e_r(G_1),\ldots,e_r(G_m) \r$, where for $j\in [1,m]_\Z$ we have
$$
e_r(G_j)=\left\{
\begin{array}{ll}
\l i,\pi_j(1) \r             &\mbox{if}\ G_j=R_{i,\cR},\\
\l 4,\pi_j(1),\pi_j(2) \r    &\mbox{if}\ G_j=M_2(N).
\end{array}\right.
$$
 Let $\bK$ be a $k$-input $m$-output $n$-bit quantum circuit
 $\bK=(K,\La_1,\La_2,S)$ based on $\cG_\cR$, 
 where $[1,n]_\Z\setminus\La_1=\{i_1,\ldots,i_{n-k}\}$ and
 $\La_2=\{j_1,\ldots,j_m\}$.
 Then the $\cG_\cR$-code of $\bK$, denoted by $c_r(\bK)$,
 is defined to be the list of finite sequences of natural numbers, 
$$
 c_r(\bK)=\l\l\braket{i_1,S(i_1)},\ldots,
\braket{i_{n-k},S(i_{n-k})}\r,c_r(K),\l j_1,\ldots,j_m\r\r.
$$
 Given a QCF $\cK=\{\bK_n\}_{n\ge 1}$ of size $s$
 based on $\cG_\cR$, the QCF $\cK$ is said to be {\em $\cG_\cR$-uniform}
 if the function $1^n\mapsto c_r(\bK_n)$ is computable 
 by a DTM in time $p(s(n))$ for some polynomial $p$.
 
The $\cG_\cR$-uniform QCFs are a subclass of
 the general uniform QCFs to be defined as follows.
 As Shor pointed out in \cite{Sho97},
 the entries of the S-matrices of quantum gates
 in a uniform QCF must be polynomial time computable numbers
\footnote{Actually, Shor \cite{Sho97} required that
 the entries should be computable in the sense that
 the first $n$ bits are computable in time polynomial in $n$,
 while we require that the first $n$ bits of a computable number
 are computable in time polynomial in $n$
 (cf.\ Bernstein-Vazirani \cite{BV97}).}.
 It follows that the entries of elementary gates must be
 restricted to be polynomially computable ones.
 Thus it is natural to assume that
 any uniform QCF can be decomposed into the elementary gates in
$$ 
\cG_{\rP\C}=\{R_{1,\th},\ R_{2,\th},\ R_{3,\th}
,\ M_2(N)|\ \th\in\rP\C\cap[0,2\pi]\}. 
$$
According to the above, we shall give the formal definition
 of uniform QCFs for QCFs based on the set $\cG_{\rP\C}$
 instead of the universal set $\cG_u$.
 For any $\th\in\rP\C$, let $c(\th)$ be the code of $\th$.
 Let $K=(G_{m},\pi_m),\ldots,(G_{1},\pi_1)$ be
 a quantum circuit based on $\cG_{\rP\C}$.
 Then the {\it code} of $K$, denoted by $c(K)$,
 is defined to be the list of finite sequences
 of natural numbers $\l e(G_1),\ldots,e(G_m) \r$,
 where for $j\in [1,m]_\Z$ we have
$$
e(G_j)=\left\{
\begin{array}{ll}
\l \l i,c(\th)\r, \pi_j(1) \r  &\mbox{if}\ G_j=R_{i,\th},\\
\l 4,\pi_j(1),\pi_j(2) \r      &\mbox{if}\ G_j=M_2(N).
\end{array}\right.
$$
 Similar to the case of the code of a quantum circuit on $\cG_\cR$,
 we can define the code of a $k$-input $m$-output
 $n$-bit quantum circuit $\bK$ based on $\cG_{\rP\C}$.  
 Given a QCF $\cK=\{\bK_n\}_{n\ge 1}$ of size $s$
 based on $\cG_{\rP\C}$, the QCF $\cK$ is said to be {\em uniform}
 if the function $1^n\mapsto c(\bK_n)$ is computable 
 by a DTM in time $p(s(n))$ for some polynomial $p$.
 It is easy to see that a $\cG_\cR$-uniform QCF is uniform.
  
As is well-known, the discrete Fourier transform
 $\ket{a}\mapsto\frac{1}{\sqrt{2^n}}
 \sum_{c=0}^{2^n-1}\mbox{exp}\left(\frac{2\pi iac}{2^n}\right)\ket{c}$, 
 where $a=0,\ldots,2^n-1,$ plays an important role in Shor's algorithm
 \cite{EJ96,Sho97}. It is easy to see that
 the polynomial size QCF $\cK=\{K_n\}_{n\ge 1}$
 that performs the discrete Fourier transform
 is such that on input $1^n$ the code of $K_n$
$$
c(K_n)=\l c^{1}(A),c^{12}(B_1),c^2(A),\ldots,c^{1n}(B_{n-1}),
\ldots,c^{(n-1)n}(B_1),c^n(A)\r 
$$\sloppy
 can be computed by a polynomial time bounded DTM,
 where $c^{j}(A)=\l \l 3,c(\pi)\r,j\r$,
 $\l \l 1,c(\pi/4)\r,j\r$ and
\begin{eqnarray*}
c^{ij}(B_k) &=&
\l \l 3,c(\pi/2^{k+1})\r,i \r,
\l \l 2,c(-\pi/2^{k+2})\r,j \r,
\l \l 3,c(\pi/2^{k+2})\r,j \r,
\l 4,i,j \r,\\
& &\l \l 2,c(\pi/2^{k+2}) \r,j\r,
\l \l 3,c(-\pi/2^{k+2}) \r,j\r,
\l 4,i,j \r,
\end{eqnarray*}
 Thus, $\cK$ is uniform.
 For example, $K_4$ is the quantum circuit illustrated
 in Figures 3 and 4.  

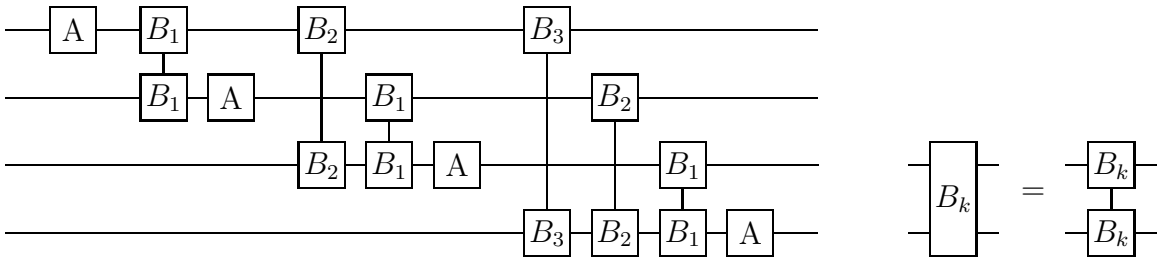
\begin{center}
\unitlength 3mm
\begin{picture}(51,15)
\put(0,3){\line(1,0){23}}\put(25,3){\line(1,0){1}}
\put(28,3){\line(1,0){1}}\put(31,3){\line(1,0){1}}
\put(34,3){\line(1,0){2}}\put(0,6){\line(1,0){13}}
\put(15,6){\line(1,0){1}}\put(18,6){\line(1,0){1}}
\put(21,6){\line(1,0){8}}\put(31,6){\line(1,0){5}}
\put(0,9){\line(1,0){6}}
\put(8,9){\line(1,0){1}}\put(11,9){\line(1,0){5}}
\put(18,9){\line(1,0){8}}\put(28,9){\line(1,0){8}}
\put(8,12){\line(1,0){5}}\put(15,12){\line(1,0){8}}
\put(25,12){\line(1,0){11}}
\put(0,12){\line(1,0){2}}\put(4,12){\line(1,0){2}}
\put(7,10){\line(0,1){1}}\put(14,7){\line(0,1){4}}
\put(30,4){\line(0,1){1}}\put(17,7){\line(0,1){1}}
\put(24,4){\line(0,1){7}}\put(27,4){\line(0,1){4}}
\put(2,11){\framebox(2,2){A}}
\put(6,8){\framebox(2,2){$B_1$}}\put(6,11){\framebox(2,2){$B_1$}}
\put(9,8){\framebox(2,2){A}}
\put(13,5){\framebox(2,2){$B_2$}}\put(13,11){\framebox(2,2){$B_2$}}
\put(26,2){\framebox(2,2){$B_2$}}\put(26,8){\framebox(2,2){$B_2$}}
\put(16,5){\framebox(2,2){$B_1$}}\put(16,8){\framebox(2,2){$B_1$}}
\put(32,2){\framebox(2,2){A}}
\put(29,2){\framebox(2,2){$B_1$}}\put(29,5){\framebox(2,2){$B_1$}}
\put(19,5){\framebox(2,2){A}}
\put(23,2){\framebox(2,2){$B_3$}}\put(23,11){\framebox(2,2){$B_3$}}
\put(40,3){\line(1,0){1}}\put(40,6){\line(1,0){1}}
\put(43,3){\line(1,0){1}}\put(43,6){\line(1,0){1}}
\put(47,3){\line(1,0){1}}\put(47,6){\line(1,0){1}}\put(50,6){\line(1,0){1}}
\put(50,3){\line(1,0){1}}\put(48,2){\framebox(2,2){$B_k$}}
\put(49,4){\line(0,1){1}}\put(48,5){\framebox(2,2){$B_k$}}
\put(41,2){\framebox(2,5){$B_k$}}\put(45,4.5){$=$}
\end{picture}
\end{center}
Figure 3\ :\ The quantum circuit $K_4$. 
In this figure, $A=R_{1,\pi/4}\cdot R_{3,\pi}$, 
 and $B_k$ is the 2-bit quantum gate
 determined by the quantum circuit $K_{B,k}$ (Figure 4) based on $\cG_u$.
 The S-matrix of $B_k$ is diag(1,1,1,exp$(\frac{i\pi}{2^k}))$,
 where diag($a_1,\ldots,a_n$) is an $n$-dimensional diagonal matrix
 whose diagonal components are $a_1,\ldots,a_n$ in this order. 

\begin{center}
\unitlength 3mm
\begin{picture}(30,7)
\put(0,2){\line(1,0){1}}\put(4,2){\line(1,0){1}}
\put(8,2){\line(1,0){1}}
\put(11,2){\line(1,0){1}}
\put(15,2){\line(1,0){1}}\put(19,2){\line(1,0){1}}
\put(22,2){\line(1,0){1}}
\put(0,5){\line(1,0){1}}
\put(4,5){\line(1,0){19}}\put(10,3){\line(0,1){2}}
\put(21,3){\line(0,1){2}}
\put(1,1){\framebox(3,2){$R_{2,k}$}}
\put(5,1){\framebox(3,2){$R_{3,k}$}}
\put(9,1){\framebox(2,2){N}}
\put(12,1){\framebox(3,2){$R_{4,k}$}}
\put(16,1){\framebox(3,2){$R_{5,k}$}}
\put(20,1){\framebox(2,2){N}}
\put(1,4){\framebox(3,2){$R_{1,k}$}}
\put(10,5){\circle*{0.5}}\put(21,5){\circle*{0.5}}
\end{picture}
\end{center}
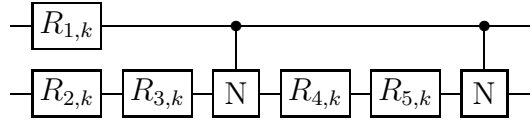
Figure 4\ :\ The quantum circuit $K_{B,k}$. In this figure,
 $R_{1,k}=R_{3,\pi/2^{k+1}},
\ R_{2,k}=R_{2,-\pi/2^{k+2}},\ \\
 R_{3,k}=R_{3,\pi/2^{k+2}},\ R_{4,k}=R_{2,\pi/2^{k+2}}$, and 
 $R_{5,k}=R_{3,-\pi/2^{k+2}}$.\\ \ 

A formal definition of a simulation of a QCF by a QTM
 is given as follows. 
 Let $M$ be a multi-track QTM such that
 the alphabet of each track contains 0 and 1.
 We say that $M$ {\em carries out an $n$-input $k$-bit quantum circuit} 
 $\bK=(K,\La_{1},\La_{2},S)$ if for every $n$-bit string $x$,
 the output state of $M$ for input state $\ket{q_0,{\rm tape}[x],0}$ is
$$
\sum_{y\in\{0,1\}^k} \ket{q_f,{\rm tape}[x,y],0}\l y|G(K)|u(x,\bK)\r.
$$
 For any function $f:\N\rightarrow\N$,
 we say that $M$ {\em simulates} a QCF
 $\cK=\{\bK_n\}_{n\ge 1}$ in time $f(n)$,
 if on every $n$-bit input string,
 $M$ carries out $\bK_n$ and
 the computation time is $f(n)$.
 Then the following lemma holds.

\begin{Lemma}\label{th:51}
For any $\cG_\cR$-uniform QCF $\cK$,
 there exist a polynomial $p$ and an SNQTM $M$
 which simulates $\cK$ in time $p(s(n))$, 
 where $s$ is the size of $\cK$.
\end{Lemma}

{\it Proof.}\ \ Let $\cK=\{\bK_n\}_{n\ge 1}$
 be a $\cG_\cR$-uniform QCF of size $s$, 
 and let $\bK_n=(K_n,\Lambda_{1,n},\Lambda_{2,n},S_n)$.
 First, we show that there exists a multi-track QTM $M$
 that carries out the following steps
 for input state $\ket{q_0,{\rm tape}[x],0}$. 
 Throughout this proof, we assume that the length of $x$ is $n$.

Step 1. Write $1^n$ on the third track,
 $c_r(\bK_n)$ on the fourth track,
 and $u=u(x,\bK_n)$ on the second track.

Step 2. Iterate the following steps 3 and 4 for $l=1$ to $s(n)$,
 where step 4 refers to step 4.1 or 4.2.

Step 3.\ \ On the fourth track, scan the $l$-th component $\l h,i\r$ 
 or $\l h,i,j\r$ of $c_r(K_{n})$, where $h\in[1,4]_\Z$ and
 $i,j\in [1,|\La_{1,n}|+|\mb{domain}(S_n)|]_\Z$.
 That is, $h$ is the index in $\cG_\cR$ of the $l$-th quantum gate
 constructing $K_n$, and $i$ (and $j$) is the bit number of the wire
 connected to the $l$-th gate.

Step 4.1.\ \ When $h=4$, if $i<j$, then
 carry out the unitary transformation
 $\ket{x,y}\mapsto\ket{x,x+y\ \mb{mod}\ 2}$
 on the cell-set $\{i,j\}$ of the second track.
 If $i>j$, then carry out the unitary transformation
 $\ket{x,y}\mapsto\ket{x+y\ \mb{mod}\ 2,y}$
 on the cell-set $\{i,j\}$ of the second track.

Step 4.2.\ \ When $h\neq 4$,
 carry out the transformation $R_{h,\cR}$
 on the cell-set $\{i\}$ of the second track. 

Step 5. Empty the fourth and the third tracks.

Since $\cK$ is a $\cG_\cR$-uniform QCF of size $s$,
 there is a DTM that computes the function $1^n\mapsto c_r(\bK_n)$
 in time polynomial of $s(n)$. Thus, we construct an SNQTM
 that carries out step 1 in time polynomial of $s(n)$
 by using the synchronization theorem,
 the addition and the permutation of tracks,
 and the dovetailing lemma.
 Moreover, using the synchronization theorem we can construct
 SNQTMs for steps 3 and 5 that run in time polynomial of $s(n)$.
 For each unitary transformation in step 4, we can construct
 an SNQTM that carries it out using the completion lemma.
 For example, an SNQTM that carries out
 the unitary transformation $R_{1,\cR}$
 is such that the quantum transition function $\de$ satisfies
$$ 
\begin{array}{ll}
\de(q_0,0,q_1,0,-1)=\de(q_0,1,q_1,1,-1)=\cos\cR,&\ \\
-\de(q_0,0,q_1,1,-1)=\de(q_0,1,q_1,0,-1)=\sin\cR,&\
 \de(q_1,B,q_{f},B,1)=1,\\
\de(q_f,a,q_0,a,1)=1\ \ (a\in\{B,0,1\}). 
\end{array}
$$
 Similarly, we can also construct SNQTMs that
 carries out the other unitary transformations. 
 Now we can construct an SNQTM that
 accomplishes step 4 by applying the addition and the permutation of tracks,
 the branching lemma, and the synchronization theorem
 to SNQTMs that carries out their unitary transformations.
 An SNQTM which carries out step 4.1 or 4.2 according to $h$
 in step 3 can be constructed by the branching lemma,
 the addition and the permutation of tracks, and the dovetailing lemma. 
 We can construct an SNQTM that carries out step 2
 by the looping lemma.
 Finally, we can construct the desired QTM $M$ by applying
 the addition and the permutation of tracks, and the dovetailing lemma
 to SNQTMs that carries out steps 1, 2, and 5.
 Each dovetailed SNQTM can be constructed
 so that the dovetailing conditions can be satisfied.

It is easy to see that $M$ carries out $K_n$
 and the computation time of $M$ is a polynomial of $s(n)$. 
From the above, $M$ simulates $\cK$ in time polynomial of $s(n)$.\ \ \ \qed

Using Theorem \ref{th:43} and Lemma \ref{th:51},
 we investigate the detailed relationships
 among complexity classes between QTMs and QCFs.
 We shall now define classes of languages efficiently
 recognized by QTMs or QCFs implementing Monte Carlo, Las Vegas,
 and exact algorithms, that is, quantum analogues of
 the probabilistic complexity classes {\bf BPP}, {\bf ZPP}, and {\bf P}. 

We say that a QTM $M$ {\em accepts} (or {\em rejects})
 $x\in\{0,1\}^*$ {\em with probability} $p$
 if the output state $\ket{\psi}$ of $M$
 for input state $\ket{q_0,{\rm tape}[x],0}$ satisfies 
$$ 
||E^{\hat{T^1}}({\rm tape}[x])E^{\hat{T^2}}({\rm tape}[1])\ket{\psi}||^2=p,
\ \ (\mb{or}\ \ ||E^{\hat{T^1}}({\rm tape}[x])
E^{\hat{T^2}}({\rm tape}[0])\ket{\psi}||^2=p).
$$
 We say that $M$ {\it recognizes} a language $L$
 with probability at least $p$
 if $M$ accepts $x$ with probability at least $p$ for any $x\in L$ 
 and rejects $x$ with probability at least $p$ for any $x\not\in L$. 
 Moreover, we say that $M$ recognizes $L$
 {\em with probability uniformly larger than} $p$,
 if there is a constant $0<\eta\le 1-p$ such that 
 $M$ recognizes $L$ with probability at least $p+\eta$. 
 Let $A$ be a subset of $\C$. 
 A language $L$ is in ${\bf BQP}_A$ (or ${\bf EQP}_A$)
 if there is a polynomial time bounded QTM $M=(Q,\Si,\de)$
 that recognizes $L$ with 
 probability uniformly larger than $\frac{1}{2}$
 (or with probability 1) and range($\de$)$\subseteq A$. 
 A language $L$ is in ${\bf ZQP}_A$
 if there is a polynomial time bounded QTM $M=(Q,\Si,\de)$
 satisfying the following conditions.

(1) $M$ recognizes $L$ with probability uniformly
 larger than $\frac{1}{2}$.

(2) range($\de$)$\subseteq A$.

(3) If $M$ accepts (rejects) input $x$ with a positive probability,
 $M$ rejects (accepts) $x$ with probability $0$.\\
 From these definitions, we have obviously
 ${\bf EQP}_A\subseteq{\bf ZQP}_A\subseteq{\bf BQP}_A$. 
 In what follows, when $A=\rP\C$, we denote ${\bf BQP}_A$, ${\bf EQP}_A$,
 and ${\bf ZQP}_A$ by ${\bf BQP}$, ${\bf EQP}$, and ${\bf ZQP}$, respectively. 

Let $M$ be an SNQTM that recognizes a language $L$
 with probability uniformly larger than $\frac{1}{2}$ in time $t(n)$,
 where $n$ is the length of the input of $M$. 
Then we can recognize $L$ with probability uniformly larger than $1-\ep$
 by iterating the computation of $M$ on the input
 $k=O(\log\frac{1}{\ep})$ times
 ($\ep$ is a positive number independent of the input)
 and calculating the majority of the $k$ answers. 
Moreover, Bennett et al.\ \cite{BBBV97}
 showed that an SNQTM that recognizes $L$
 with probability uniformly larger than $1-\ep$ in time $ct(n)$
 (here, $c$ is a polynomial in $\log\frac{1}{\ep}$ and independent of $n$)
 can be constructed. This fact means that the classes {\bf BQP} and
 {\bf ZQP} we have now defined are identical with {\bf BQP} and
 {\bf ZQP} defined in \cite{BV97,BB94}.
 
A definition of recognition of languages
 by quantum circuits is given as follows.
 Let $K$ be an $n$-input 2-output quantum circuit and $x\in\{0,1\}^n$.
 When $\rh^K(01|x)=p$ (or $\rh^K(00|x)=p$),
 we say that $K$ {\em accepts} (or {\em rejects}) $x$
 {\em with probability} $p$.  
 For any language $L_n\subseteq\{0,1\}^n$,
 we say that $K$ {\em recognizes} $L_n$ with probability at least $p$ 
 if $K$ accepts $x$ with probability at least $p$
 for any $x\in L_n$ and $K$ rejects $x$ with probability
 at least $p$ for any $x\not\in L_n$. 

We need to consider circuit families in order to recognize
 languages including strings with different lengths. 
 In what follows, we write $L_n=L\cap\{0,1\}^n$ for any language $L$. 
 We say that a {\em QCF} $\cK=\{K_n\}_{n\ge 1}$ {\em recognizes}
 a language $L$ {\em with probability} at least $p$ 
 if $K_n$ recognizes $L_n$ with probability at least $p$ for any $n\in\N$. 
 We say that $\cK$ recognizes a language $L$
 {\em with probability uniformly larger than} $p$
 if there is a constant $0<\eta\le 1-p$
 such that $K_n$ recognizes $L_n$
 with probability at least $p+\eta$ for any $n$.
 We say that a language $L$ has
 {\em bounded-error (or exact) uniform polynomial size quantum circuits},
 in symbols $L\in{\bf BUPQC}$ (or $L\in{\bf EUPQC}$),
 if there is a uniform polynomial size QCF $\cK=\{K_n\}_{n\ge 1}$ 
 that recognizes $L$ with probability uniformly larger than $\frac{1}{2}$\ 
 (with probability 1).
 Moreover, we say that a language $L$ {\em has zero-error uniform 
 polynomial size quantum circuits}, in symbols $L\in{\bf ZUPQC}$,
 if there is a uniform polynomial size QCF $\cK=\{K_n\}_{n\ge 1}$
 recognizing with probability uniformly larger than $\frac{1}{2}$
 and satisfying 
 $\rh^{K_{|x|}}(00|x)=0$ or $\rh^{K_{|x|}}(01|x)=0$
 for any $x\in\{0,1\}^*$. 
 From these definitions we have obviously
 ${\bf EUPQC}\subseteq {\bf ZUPQC}\subseteq {\bf BUPQC}$. 

As is well-known, $\bP$ is identical with
 the class of languages that have
 uniform polynomial size Boolean circuits\footnote{
In this paper, uniform Boolean circuit families mean 
 polynomial time uniform ones. In computational complexity theory,
 more restricted families have been investigated
 and some of them are also equivalent to polynomial time bounded DTMs.}
 \cite{Pap94}.
 The following identical relation holds between complexity classes of
 QTMs and QCFs. This relation means that
 QTMs and uniform QCFs are equivalent as probabilistic machines
 implementing Monte Carlo algorithms as suggested by Shor \cite{Sho97}.

\begin{Theorem}\label{th:52}
 ${\bf BQP}={\bf BUPQC}$.
\end{Theorem}

{\it Proof.}\ Let $L\in{\bf BQP}$. Then without loss of generality,
 we can assume that there is a QTM $M=(Q,\Si,\de)$
 that recognizes $L$ with probability
 uniformly larger than $\frac{1}{2}$ in time $p(n)$,
 where $p$ denotes a polynomial (See Remark 2).
 This QTM $M$ can be $p(n)$-simulated by a quantum circuit $K_{n}$ 
 of size $O(p^2(n))$ constructed as the proof of Theorem \ref{th:43}.
 The quantum gate $G(K_n)$ can be decomposed into
 two sorts of quantum gates $G_1$ and $G_2$
 as given in the proof of Theorem \ref{th:43},
 and the array of $G_1$ and $G_2$ in $K_n$
 can be computed in time polynomial in $n$. Moreover,   
 range($\de$)$\subseteq \rP\C$ by the definition of ${\bf BQP}$,
 so that from the S-matrix of $G_1$ we can compute
 the array of elementary gates in $\cG_{\rP\C}$
 decomposing $G_1$ in time independent of $n$.
 Obviously, $G_2$ can be decomposed into elementary gates in $\cG_{\rP\C}$
 in time polynomial in $n$. Therefore, there is a DTM which on input $1^n$
 produces the code $c(K_n)$ in time polynomial in $n$.
 Thus, $\cK=\{K_n\}_{n\ge 1}$ is uniform. 
 From the above, $L\in {\bf BUPQC}$.

Conversely, suppose $L\in{\bf BUPQC}$. Then, for all $n\in\bN$
 there is a quantum circuit $K_n$ of size $p(n)$ based on $\cG_{\rP\C}$
 which recognizes $L_n$ with probability $\frac{1}{2}+\et$,
 where $p$ is a polynomial and
 $0<\eta\le\frac{1}{2}$ is a constant independent of $n$. 
 Moreover, there is a DTM $M_0$ that computes
 the function $1^n\mapsto c(K_n)$ in time polynomial in $n$.
 Assume that the length of a bit string $x$ is $n$.
 Let $c(K_n)=\l e(G_1),\ldots,e(G_k),\ldots,e(G_{p(n)})\r$,
 where $e(G_k)=\l \l i,c(\th)\r,\pi_k(1)\r$ if $G_k=R_{i,\th}$
 and $e(G_k)=\l 4,\pi_k(1),\pi_k(2)\r$ if $G_k=M_2(N)$. 
 Now we compute the $\cG_\cR$-code $c_r(K_{n,\ep})$
 of a quantum circuit $K_{n,\ep}$ based on $\cG_\cR$
 such that $||G(K_n)-G(K_{n,\ep})||\le\ep$ from $c(K_n)$ as follows.
 For each $k=1,\ldots,p(n)$, from the component $\l i,c(\th)\r$
 of $c(K_n)$ representing $G_k=R_{i,\th}$ in $K_n$,
 we compute an integer $m$ such that $||R_{i,\th}-R_{i,\cR}^m||
\le\frac{\ep}{p(n)}$ by Lemma \ref{th:42}, and
 replace the component $\l\l i,c(\th)\r,\pi_k(1)\r$ in $c(K_n)$
 by $\underbrace{\l i,\pi_k(1)\r,\ldots,\l i,\pi_k(1)\r}_m$.
 It is easy to see that the computation time of
 this algorithm is at most a polynomial in $n$ and $\log\frac{1}{\ep}$.
 Now let $\ep\le\frac{\et}{2}$.
 Then the QCF $\cK_\ep=\{K_{n,\ep}\}_{n\ge 1}$
 based on $\cG_\cR$ recognizes $L$ with probability 
 at least $\frac{1}{2}+\frac{\et}{2}$. 
 Next, we consider the $\cG_\cR$-size of $K_{n,\ep}$. 
 For each 1-bit quantum gate $R_{j,\th}$ ($j=1,2,3,\ \th\in[0,2\pi]$)
 constructing $K_n$,
 the positive integer $m$ determined by Lemma \ref{th:42}
 such that $||R_{j,\th}-R_{j,\cR}^m||\le\frac{\ep}{p(n)}$
 is at most $O(p^4(n)/\ep^4)$.
 Thus the $\cG_\cR$-size $s(n)$ of $K_{n,\ep}$ is at most
 $s(n)=O(p^4(n)/(\frac{\et}{2})^4)\times p(n)=O(p^5(n))$. 
 Therefore, $\cK_\ep$ is a $\cG_\cR$-uniform QCF of size $s(n)$.
 Applying Lemma \ref{th:51} to $\cK_\ep$, given as input
 an $n$-bit string there is a QTM $M=(Q,\Si,\de)$ that
 carries out $K_{n,\ep}$ in time $O(q(s(n)))$, where
 $q$ is a polynomial. From the proof of Lemma \ref{th:51},
 it is easy to see that range($\de$)$\subseteq \rP\C$. 
 Therefore we conclude $L\in{\bf BQP}$.\ \ \ \qed

{\em Remark 1.} Using the proof of ${\bf BUPQC}\subseteq{\bf BQP}$
 in Theorem \ref{th:52}, we can show the existence of
 a polynomial time bounded universal QTM which simulates
 any given uniform QCF with any accuracy.   

{\em Remark 2.} Any polynomial time bounded QTM $M$
 can be simulated by a two-tape QTM $M'$ whose computation time is
 exactly a polynomial in the length of the input,
 using time constructible functions to count the number of steps,
 as follows:
 (1) $M'$ writes $1^{p(n)}$ on the second tape,
 where $n$ is the length of the input,
 $p(n)$ is a time constructible polynomial,
 and the computation time of $M$ is bounded by $p(n)$;
 (2) In every time when $M'$ carries out one step of $M$ on the first tape,
 $M'$ changes $1$ to $B$ on the second tape;
 (3) When $M'$ completes the computation of $M$, the first tape
 of $M$ does not change the contents of the first tape any more,
 while $M'$ changes $1$ to $B$ on the second tape;
 (4) if the second tape scans $B$, then $M'$ halts.     
 
 The following theorem can be verified by a proof
 similar to that of ${\bf BQP}\subseteq{\bf BUPQC}$
 in Theorem \ref{th:52}, and means that QTMs
 are not more powerful than uniform QCFs
 as probabilistic machines implementing 
 exact or Las Vegas algorithms.

\begin{Theorem}\label{th:53}
{\rm (1)} ${\bf EQP}\subseteq {\bf EUPQC}$.

{\rm (2)} ${\bf ZQP}\subseteq {\bf ZUPQC}$.
\end{Theorem}
 
It is open whether the inclusion relations in Theorem \ref{th:53}
 are proper or not.
 In the proof of ${\bf BUPQC}\subseteq{\bf BQP}$ in Theorem \ref{th:52},
 we are allowed to replace quantum gates with some additional errors,
 while an analogous argument does not work in Theorem \ref{th:53}.
 
It has been considered that Shor's factoring algorithm
 is a Las Vegas quantum algorithm. 
 We shall show this fact by proving
 that a certain language corresponding to
 the factoring problem is not only in ${\bf ZUPQC}$
 but also in ${\bf ZQP}$.
 The factoring problem is polynomial time Turing reducible to
 the language ${\rm FACTOR}=
\{\l N,k\r|\ N$ has a non-trivial prime factor larger than $k\ \}$
 and the class of problems solved by Las Vegas algorithms
 is closed under polynomial time Turing reductions. On the other hand, 
 as suggested by Theorem \ref{th:53}, any language in ${\bf ZQP}$
 can be recognized most typically by a Las Vegas quantum algorithm.
 Thus, in order to verify that
 Shor's factoring algorithm is a Las Vegas quantum algorithm,
 it is sufficient to show that ${\rm FACTOR}$ is in ${\bf ZQP}$.

\begin{Theorem}\label{th:54}
 ${\rm FACTOR}\in{\bf ZQP}$.
\end{Theorem}

{\it Proof.}
 Let $\l N,k\r$ be an input of the algorithm to be constructed.
 In the following algorithm that recognizes ${\rm FACTOR}$,
 we use a Las Vegas primality testing algorithm 
 (for example, such an algorithm can be constructed
 by the algorithm of Solovay and Strassen \cite{SS77}
 and the algorithm of Adleman and Huang \cite{AH92})
 and Shor's factoring algorithm \cite{Sho97}.
 At first, let $\mb{LIST}=\{N\}$.

Step 1. Carry out steps 2--4
 while the greatest number in LIST is larger than $1$.

Step 2. For the greatest number $N'$ in LIST,
 check whether $N'$ is prime or not
 by the Las Vegas primality testing algorithm.
 If $N'$ is judged to be prime, then go to step 3.
 If $N'$ is judged to be composite, go to step 4.
 Otherwise, output a special mark `?' and end.

Step 3. If $N'>k$ then output $1$ and end.
 Otherwise, output $0$ and end.

Step 4. On input $N'$, carry out Shor's factoring algorithm.
 If a factor $p$ is found, then replace $N'$ in LIST by $p$ and $N'/p$,
 and go to step 2.
 If no factor is found, output `?' and end. 

Step 2 can be implemented by a polynomial time bounded SNQTM,
 because {\bf ZPP} is included in {\bf ZQP}.\ Step 3 can
 also be implemented by a polynomial time bounded SNQTM
 using the synchronization theorem. In step 4   
 we can divide Shor's factoring algorithm into three processes:
 (1) a process that produces a factor candidate of $N'$;
 (2) a process that iterates process (1) $j=O((\log N)^2)$ times
 in order to obtain $j$ factor candidates;
 (3) a process that produces a true factor
 if the factor exists in the $j$ candidates,
 and otherwise produces `?'.
 Note that process (1) also includes
 a deterministic algorithm performed efficiently
 for the case where $N'$ is an even number or a prime power.
 We have shown that the discrete Fourier transform
 can be done by a uniform polynomial size QCF in this section. 
 Using a similar way, we can make sure that
 process (1) can be carried out by a uniform polynomial size QCF $\cK$.
 Let $\ep>0$ be a small constant independent of $N$.
 Similar to the proof of Theorem \ref{th:52}, for any $K_n$ in $\cK$,
 the $\cG_\cR$-code of a quantum circuit $K_{n,\ep}$
 based on $\cG_\cR$ such that $||G(K_n)-G(K_{n,\ep})||\le\ep$
 can be computed in time polynomial in $n$.
 Thus $\cK_\ep=\{K_{n,\ep}\}_{n\ge 1}$ is $\cG_\cR$-uniform.
 We can construct an SNQTM $M_1$ that carries out
 $\cK_\ep$ by Lemma \ref{th:51}.
 An SNQTM $M_2$ which carries out process (2)
 can be constructed by inserting $M_1$
 into a looping machine $j$ times.
 We can construct an SNQTM $M_3$ 
 that carries out process (3) by the synchronization theorem, 
 and construct an SNQTM $M$ implementing
 Shor's factoring algorithm by applying the addition and the permutation
 of tracks and the dovetailing lemma to $M_2$ and $M_3$. 

In step 4 the probability that produces `?' is less than $1/N$,
 since by one round of process (1) we get a true factor
 with probability at least $\Omega(1/\log N)$
 and we repeat process (1) $O((\log N)^2)$ rounds
 to reduce the probability that produces `?' up to less than $1/N$. 
 In step 2, by iterating the Las Vegas primality testing
 a polynomial number of times we can make
 the probability that produces `?' less than $1/N$. 
 Moreover, steps 2--4 will be carried out at most $\log N$ times.
 Thus the above algorithm produces `?'
 with probability at most $\eta<1/2$,
 where $\eta$ is independent of the input. 
 Now it is easy to conclude that ${\rm FACTOR}\in {\bf ZQP}$.\ \ \ \qed

{\it Remark.} From Theorems \ref{th:53} and \ref{th:54} 
 it follows that ${\rm FACTOR}\in{\bf ZUPQC}$. 
 However, this fact can be verified in a more
 straightforward argument. In fact, we have verified
 that Shor's factoring algorithm (step 4)
 in the algorithm of the proof of Theorem \ref{th:54}
 can be implemented by a uniform polynomial size QCF.
 On the other hand, the other part of the algorithm can be written
 as a classical probabilistic algorithm.
 Coin flips can be implemented by Hadamard gates,
 and the classical deterministic part can
 be implemented by Toffoli gates. 
 These two sorts of gates can be decomposed
 into $O(1)$ elementary gates in $\cG_{\rP\C}$.   
 Thus, the other part of the algorithm can be also
 implemented by a uniform polynomial size QCF.

By analogous arguments, we can also show that Shor's algorithm
 for the discrete logarithm problem defined in \cite{Sho97}
 is a Las Vegas quantum algorithm.

Considering the proof of Theorem \ref{th:54}, it might be expected
 that {\bf ZQP} is equal to {\bf ZUPQC}. 
 However, we should notice that the above algorithm uses
 a Las Vegas type primality testing to produce a correct answer.
 This primality testing prevents us from producing incorrect answers.
 But this check-algorithm is classical Las Vegas one.
 A classical Las Vegas algorithm can be exactly carried out
 by a Las Vegas type QTM, since a polynomial time bounded
 probabilistic Turing machine can be exactly simulated
 by a polynomial time bounded QTM.
 Now, in the case where such a check-algorithm is carried out
 by a uniform QCF, it is not known whether
 we can implement this algorithm by a QTM.
 Thus, even if a quantum algorithm is carried out efficiently
 by a Las Vegas type uniform QCF,
 we cannot say that the algorithm is efficiently carried out
 by a Las Vegas type QTM.

The state transition of a QTM is determined
 by the quantum transition function,
 finite numbers of complex numbers, while
 in order to characterize that of a QCF,
 we can use infinite numbers of complex numbers  
 even under the uniformity condition.
 This suggests that some QCF cannot be simulated exactly by a QTM.
 In fact, we can show that a QCF carrying out the discrete Fourier transform
 cannot be {\em exactly} simulated by any QTM as follows.

\begin{Proposition}\label{th:545}
 A QCF $\cK=\{K_n\}_{n\ge 1}$ carrying out the discrete Fourier transform
$$
\ket{a}\mapsto\frac{1}{\sqrt{2^n}}\sum_{c=0}^{2^n-1}
{\rm exp}\left(\frac{2\pi iac}{2^n}\right)\ket{c},
$$ 
 where $a=0,\ldots,2^n-1$, cannot be exactly simulated
 by any QTM.
\end{Proposition}

{\it Proof.} Let $\overline{\Q}$ be the algebraic closure
 of $\Q$. Let $F(\al_1,\ldots,\al_m)$ be the field generated
 by $\al_1,\ldots,\al_m$ over a field $F$.
 The range of the quantum transition function
 of a QTM $M=(Q,\Si,\de)$ consists of finite numbers of complex numbers
 $\{\al_1,\ldots,\al_m\}$. Thus,
 the set $\{\l C'|M_\de^t|C\r\ |\ C',C\in\cC(Q,\Si),\ t\in\Z_{\ge 0}\}$
 is included in an extended field $\Q(\al_1,\ldots,\al_m)$ of $\Q$.
 On the other hand, the $2^n$-dimensional unitary matrix
 representing the quantum gate $G(K_n)$
 determined by $K_n$ contains the complex number $e^\frac{2\pi i}{2^n}$
 as the components.
 Therefore, it is sufficient to show the relation
 $\{e^\frac{2\pi i}{2^n}\ |\ n\in\N\}
 \not\subseteq\Q(\al_1,\ldots,\al_m)$.
 The dimension of the vector space
 $\Q(e^{\frac{2\pi i}{2}},\ldots,e^{\frac{2\pi i}{2^n}})
=\Q(e^{\frac{2\pi i}{2^n}})$ over $\Q$ is $2^{n-1}$.
 Moreover, $\Q(e^{\frac{2\pi i}{2^n}})\subseteq \overline{\Q}$. 
 Henceforth, let $F_k=\Q(\al_1,\ldots,\al_k)\cap\overline{\Q}$. 
Now, we shall show that $F_k$ is a finite extension of $\Q$
 by induction on $k$. When $k=0$, it is trivial.
 Suppose that $F_k$ is a finite extension of $\Q$.
 If $F_{k+1}=F_k$, then it is easy to see that
 $F_{k+1}$ is a finite extension of $\Q$.
 Now, suppose that $F_{k+1}\neq F_k$ and
 let $\ga\in F_{k+1}\setminus F_k$.
 Then there is a non-constant rational expression $f(x)$ over
 $\Q(\al_1,\ldots,\al_k)$ such that $\ga=f(\al_{k+1})$.
 Since $\ga$ is in $\overline{\Q}\setminus\Q$, there is
 a minimal polynomial $g$ over $\Q$ of $\ga$,
 so that we have $g\circ f(\al_{k+1})=g(\ga)=0$.
 It follows that $\al_{k+1}$ is algebraic over $\Q(\al_1,\ldots,\al_k)$.
 Supposing that $l$ is the dimension of the vector space
 $\Q(\al_1,\ldots,\al_{k+1})$ over $\Q(\al_1,\ldots,\al_k)$,
 the degree of $\ga$ over $\Q(\al_1,\ldots,\al_k)$ is at most $l$.
 Let $h_1$ be the minimal polynomial over $\Q(\al_1,\ldots,\al_k)$
 of $\ga$. Since $\ga$ is also algebraic over $\Q$ and
 $\Q\subseteq\Q(\al_1,\ldots,\al_k)$, the polynomial $h_1$ divides
 the minimal polynomial $h_2$ over $\Q$ of $\ga$. 
 The coefficients of $h_1$ are in $\overline{\Q}$, since
 $h_2$ is uniquely decomposable over $\overline{\Q}$.
 Thus, the coefficients of $h_1$ are in $F_k$,
 so that the degree of $\ga$ over $F_k$ is at most $l$.
 Therefore, $F_{k+1}$ is a finite extension
 of $F_k$. By inductive hypothesis,
 $F_{k+1}$ is a finite extension of $\Q$.
 Therefore, $\Q(\al_1,\ldots,\al_m)\cap\overline{\Q}$
 is a finite extension of $\Q$,
 and hence we have $\{e^\frac{2\pi i}{2^n}\ |\ n\in\N\}
 \not\subseteq\Q(\al_1,\ldots,\al_m)$.\ \ \ \qed 

Thus, there is a fair chance that ${\bf EQP}\neq{\bf EUPQC}$
 or that ${\bf ZQP}\neq{\bf ZUPQC}$. 
      
Next we introduce the notion of the uniformity of QCFs
 based on finite subsets of $\cG_u$
 and consider classes of languages recognized by such QCFs.

Assume that a finite set $\cG$ of quantum gates
 is indexed as $\cG=\{G_1,\ldots,G_l\}$,
 where $G_i$ is an $n_i$-bit quantum gate 
 for $i=1,\ldots,l$. 
 Let $K=(G_{i_m},\pi_m),\ldots,(G_{i_1},\pi_1)$ be
 a quantum circuit based on $\cG$. 
Then the {\it $\cG$-code} $c_\cG(K)$
 is defined to be the list of finite sequences of natural numbers, 
 $\l\l i_1,\pi_1(1),\pi_1(2),\ldots,\pi_1(n_{i_1})\r,\ldots,
 \l i_m,\pi_m(1),\pi_m(2),\ldots,\pi_m(n_{i_m})\r\r.$ 
Moreover, let $\bK$ be a $k$-input $m$-output $n$-bit quantum circuit
 $\bK=(K,\La_1,\La_2,S)$ based on $\cG$, where
 $[1,n]_\Z\setminus\La_1=\{i_1,\ldots,i_{n-k}\}$ and
 $\La_2=\{j_1,\ldots,j_m\}$. 
 Then the $\cG$-code of $\bK$, denoted by $c_\cG(\bK)$, is defined by
 the list of finite sequences of natural numbers, 
$$ 
c_\cG(\bK)=\l\l\braket{i_1,S(i_1)},\ldots,\braket{i_{n-k},S(i_{n-k})}\r,
c_\cG(K),\l j_1,\ldots,j_m\r\r.
$$
A QCF $\cK=\{\bK_n\}_{n\ge 1}$ of size $s$ based on $\cG$ 
 is said to be {\em $\cG$-uniform}
 if the function $1^n\mapsto c_\cG(\bK_n)$ is computable 
 by a DTM in time $p(s(n))$ for some polynomial $p$. 
 Furthermore, a QCF $\cK$ is said to be
 {\em semi-uniform}
 if there is a finite set $\cG\subseteq\cG_u$
 such that $\cK$ is $\cG$-uniform. 
 Now the following lemma holds
 similar to Lemma \ref{th:51}.

\begin{Lemma}\label{th:55}
For any semi-uniform QCF $\cK$, 
 there exist a polynomial $p$ and a QTM $M$
 which simulates $\cK$ in time $p(s(n))$,
 where $s$ is the size of $\cK$.
\end{Lemma}

We say that a language $L$ has
 {\em bounded-error (or exact)
 semi-uniform polynomial size quantum circuits}, 
 if there is a semi-uniform polynomial size QCF
 $\cK=\{K_n\}_{n\ge 1}$ that recognizes $L$
 with probability uniformly larger than $\frac{1}{2}$ (with probability 1). 
 In this case, we write $L\in{\bf BSPQC}$ (or $L\in{\bf ESPQC}$). 
 We say that $L$ has 
 {\em zero-error semi-uniform polynomial size quantum circuits},
 if there is a semi-uniform polynomial size QCF
 $\cK=\{K_n\}_{n\ge 1}$ recognizing $L$
 with probability uniformly larger than $\frac{1}{2}$ and satisfying 
 $\rh^{K_{|x|}}(00|x)=0$ or $\rh^{K_{|x|}}(01|x)=0$ for any $x\in\{0,1\}^*$. 
 In this case, we write $L\in{\bf ZSPQC}$. 
 From these definitions we have obviously
 ${\bf ESPQC}\subseteq{\bf ZSPQC}\subseteq{\bf BSPQC}$. 

The following theorem shows that semi-uniform polynomial size QCFs
 are equivalent to polynomial time bounded QTMs
 whose transition amplitudes are arbitrary complex numbers.

\begin{Theorem}\label{th:56}
{\rm (1)} ${\bf BQP}_\C={\bf BSPQC}$.

{\rm (2)} ${\bf EQP}_\C={\bf ESPQC}$.

{\rm (3)} ${\bf ZQP}_\C={\bf ZSPQC}$.
\end{Theorem}

{\em Proof}. We shall show only statement (1). 
 Statements (2) and (3) can be proved similarly. 

Let $L\in {\bf BQP}_\C$.
 Then, there is a QTM $M=(Q,\Si,\de)$ that recognizes $L$
 with probability uniformly larger than $\frac{1}{2}$
 in time $p(n)$, where $p$ denotes a polynomial.
 For any $n\in\N$ there is a quantum circuit $K_n$
 of size $O(p^2(n))$ that $p(n)$-simulates $M$
 by Theorem \ref{th:43}.
 We use the same notations as the proof of Theorem \ref{th:43}
 by identifying $K_n$ with $K$ in this proof.
 Then the quantum gates $G_1$ and $G_2$ constructing $K_n$
 are decomposable by at most $q(n)$ gates in a finite subset $\cG$ of $\cG_u$,
 where $q(n)$ is a polynomial.
 If $\cG$ is indexed, there is a DTM that computes
 the function $1^n\mapsto c_\cG(K_n)$ in time polynomial in $n$
 by the construction of the quantum circuit in Theorem \ref{th:43}. 
 Thus, $\cK=\{K_n\}_{n\ge 1}$ is a semi-uniform polynomial size QCF
 that recognizes $L$ with probability uniformly larger than $\frac{1}{2}$.

Conversely, suppose $L\in{\bf BSPQC}$.
 Then, there is a semi-uniform polynomial size QCF $\cK=\{K_n\}_{n\ge 1}$ 
 that recognizes $L$ with probability uniformly larger than $\frac{1}{2}$. 
By Lemma \ref{th:55}, given as input an $n$-bit string,
 there is a QTM $M$ that carries out $K_n$ in time $O(p(n))$,
 where $p$ is a polynomial. 
Thus, $M$ recognizes $L$ with probability
 uniformly larger than $\frac{1}{2}$.\ \ \ \qed

{\em Remark.}\ Unlike Theorem \ref{th:52},
 the proof of the existence of a quantum circuit that
 recognizes $L\in {\bf BQP}_\C$ in Theorem \ref{th:56}
 is non-constructive.  
 For example, if a language $L$ can be recognized
 with probability uniformly larger than $\frac{1}{2}$
 by a polynomial time bounded QTM $M$, 
 there is a semi-uniform polynomial size QCF
 $\cK$ that recognizes $L$
 with probability uniformly larger than $\frac{1}{2}$, 
 but we do not know
 how to find out $\cK$ from $M$ efficiently.

Similar to the proof of Theorem \ref{th:56},
 by modifying the proof of Theorem \ref{th:44} non-constructively,
 we can show that SNQTMs (and QTMs with the binary tapes
 by Lemma \ref{th:32}) are equivalent to multi-tape QTMs
 as machines implementing not only Monte Carlo algorithms
 but exact ones from the viewpoint of the polynomial time complexity.

\begin{Theorem}\label{th:57}
For any QTM $M$, there is an SNQTM $M'$ (depending on $M$)
 that given any positive integer $t$,
 simulates $M$ for $t$ steps.
 Moreover, if $M$ is in $\rP\C$, then so is $M'$. 
\end{Theorem}

Adleman, DeMarrais, and Huang \cite{ADH97} have shown that
 if all complex numbers are allowed as transition amplitudes of QTMs,
 for any language $L$, there exists a language $L'\in{\bf BQP}_\C$
 which is Turing equivalent to $L$. 
 As a result, {\bf BSPQC} is also a set with uncountable cardinality.

Figure 5 summarizes the inclusions among the classes of languages
 which we have discussed in this section.

\begin{center}
\unitlength 3mm
\begin{picture}(42,24)
\put(14,2){{\bf P}}
\put(14,6){{\bf EQP}}\put(14,10){${\bf EUPQC}$}
\put(2,10){${\bf ESPQC}={\bf EQP}_\C$}
\put(8,14){${\bf ZSPQC}={\bf ZQP}_\C$}
\put(6,9.5){\line(3,-1){8}}
\put(6,11.5){\line(2,1){4}}
\put(12.5,15.5){\line(2,1){12}}
\put(24,6){{\bf ZPP}}
\put(23.5,11.5){\line(-3,1){6}}
\put(24,10){{\bf ZQP}}\put(24,14){${\bf ZUPQC}$}
\put(34,10){{\bf BPP}}
\put(30,18){${\bf BUPQC}={\bf BQP}$}
\put(21,22){${\bf BSPQC}={\bf BQP}_\C$}
\put(15,3.5){\line(0,1){2}}
\put(15,7.5){\line(0,1){2}}
\put(25,7.5){\line(0,1){2}}\put(25,11.5){\line(0,1){2}}
\put(35,11.5){\line(0,1){6}}
\put(16,3.5){\line(3,1){8}}
\put(26,7.5){\line(3,1){8}}
\put(24,9.5){\line(-3,-1){7}}
\put(16,11.5){\line(3,1){8}}
\put(26,15.3){\line(3,1){7}}
\put(25,21.5){\line(3,-1){7}}
\end{picture}

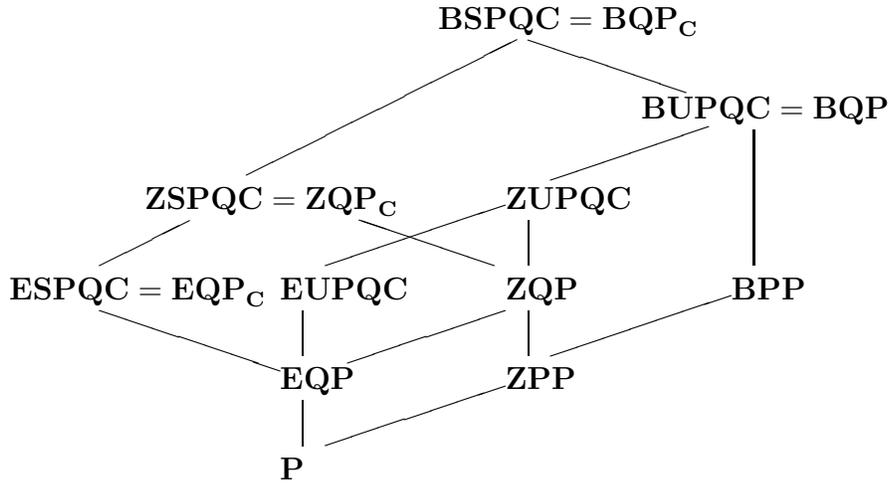
Figure 5 : The inclusions among the classes of languages
 discussed in this section.\\

\ \\
\end{center}
{\bf Appendix A. The generalization of the construction of Theorem \ref{th:43}
 to multi-tape QTMs}\\

\noindent
We can extend the construction of Theorem \ref{th:43}
 to multi-tape QTMs. In what follows, let $\vec{a}=(a_1,\ldots,a_k)$,
 $\vec{a_j}=(a_{j1},\ldots,a_{jk})$, and $\Si=\Si_1\times\cdots\times\Si_k$.
 Let $M=(Q,\Si,\de)$ be a $k$-tape QTM. This time we use
 $l_0+\sum_{j=1}^k(2t+1)(2+\lceil \log|\Si_j|\rceil)$
 wires for the simulation.
 Conditions (i) and (ii) in the proof of Theorem \ref{th:43}
 are modified as follows; we denote
$$
\ket{q;\si_{11}\bar{2};\si_{21}\bar{0};\si_{31}\bar{0};
\cdots;\si_{1k}\bar{2};\si_{2k}\bar{0};\si_{3k}\bar{0}},\ldots,
\ket{q;\si_{11}\bar{0};\si_{21}\bar{0};\si_{31}\bar{2};
\cdots;\si_{1k}\bar{0};\si_{2k}\bar{0};\si_{3k}\bar{2}}\ \ \mb{by}
$$ 
$$
\ket{q;\si_{11}\si_{21}\si_{31};\cdots;\si_{1k}\si_{2k}\si_{3k};-1,\ldots,-1},
\ldots,
\ket{q;\si_{11}\si_{21}\si_{31};\cdots;\si_{1k}\si_{2k}\si_{3k};1,\ldots,1}
$$
 respectively.

(i') $G_1\ket{w_{p,\vec{\si_1},\vec{\si},\vec{\si_3}}}
=\ket{v_{p,\vec{\si_1},\vec{\si},\vec{\si_3}}}$, where 
\begin{eqnarray*}
\ket{w_{p,\vec{\si_1},\vec{\si},\vec{\si_3}}}
 &=&\ket{p;\si_{11}\bar{0};\si_1\bar{1};\si_{31}\bar{0};\cdots;
    \si_{1k}\bar{0};\si_2\bar{1};\si_{3k}\bar{0}},\\
\ket{v_{p,\vec{\si_1},\vec{\si},\vec{\si_3}}}
 &=& \sum_{q,\vec{\ta},\vec{d}}
\de(p,\vec{\si},q,\vec{\ta},\vec{d})
 \ket{q;\si_{11}\ta_1\si_{31};\cdots;\si_{1k}\ta_k\si_{3k};\vec{d}}
\end{eqnarray*}
 for any $(p,\vec{\si_1},\vec{\si},\vec{\si_3})
\in Q\times\Si^3$; 
the summation $\sum_{q,\vec{\ta},\vec{d}}$
 is taken over all $(q,\vec{\ta},\vec{d})
\in Q\times\Si\times[-1,1]_\Z^k$.

(ii') $G_1\ket{h}=\ket{h}$ for each vector $\ket{h}$ in
 the subspace $H$ of $\C^{2^{l_0+3l}}$ spanned by
$$ 1+2\times\sum_{j=0}^{k-1}5^j=\frac{1}{2}(2^k+1) $$
types of vectors, where $l=\sum_{j=1}^k(2t+1)(2+\lceil \log|\Si_k|\rceil)$.

(1) $\ket{q;\si_{11}s_{11};\si_{21}s_{21};\si_{31}s_{31}
;\cdots;\si_{1k}s_{1k};\si_{2k}s_{2k};\si_{3k}s_{3k}}$,\\
where $\vec{s_2}\neq (\bar{1},\ldots,\bar{1})$
 and none of $s_{1i},s_{2i},s_{3i}$ are equal to
 $\bar{2}$ for some $i\in[1,k]_\Z$.

(2) For each $j\in[1,k]_\Z$ and $(D_{k-j+1},\ldots,D_k)\in
[1,2]_\Z\times[-2,2]_\Z^{j-1}$, we have
\begin{eqnarray*}
\lefteqn{
\ket{u^{j,D_{k-j+1},\ldots,D_k}_{
p,\si_{11},\si_1,\si_{31},\ldots,\si_{1(k-j)},\si_{k-j},\si_{3(k-j)},
h(D_{k-j+1}),\ldots,h(D_k)}}}\quad\\
\ \ \ \ \ \ \ \ =\sum&[&\!\!\!\! \de(p,\vec{\si},q,\vec{\ta},\vec{d})
\ket{q;\si_{11}\ta_1\si_{31};\cdots;\si_{1(k-j)}\ta_{k-j}\si_{3(k-j)}}\\
& &\otimes\ket{f(D_{k-j+1});\cdots;f(D_k);
d_1,\ldots,d_{k-j},g(D_{k-j+1}),\ldots,g(D_k)}\ ],
\end{eqnarray*}
where the summation is taken over $q\in Q$, $\ta_m\in\Si_m$,
 $d_m\in[-1,1]_\Z$ for $m\in[1,k-j]_\Z$, and
 $\ta_n\in S(D_n)$, $d_n\in S'(D_n)$ for $n\in[k-j+1,k]$.
 Here, for $i\in[k-j+1,k]_\Z$, we have
$$ 
h(D_i)=\left\{
\begin{array}{ll}
\si_i,\ta_i,\si_{1i},\si_{2i},\si_{3i}&\mbox{if}\ D_i=\pm 2,\\
\si_{1i},\si_{2i},\si_{i}             &\mbox{if}\ D_i=-1,\\
\si_{1i},\si_{i},\si_{3i}             &\mbox{if}\ D_i=0,\\
\si_{i},\si_{2i},\si_{3i}             &\mbox{if}\ D_i=1,
\end{array}\right.
f(D_i)=\left\{
\begin{array}{ll}
\si_{1i}\si_{2i}\si_{3i}              &\mbox{if}\ D_i=\pm 2,\\
\si_{1i}\si_{2i}\ta_{i}             &\mbox{if}\ D_i=-1,\\
\si_{1i}\ta_{i}\si_{3i}             &\mbox{if}\ D_i=0,\\
\ta_{i}\si_{2i}\si_{3i}             &\mbox{if}\ D_i=1,
\end{array}\right.
$$  
 $$  
g(D_i)=\left\{
\begin{array}{ll}
\mp 1              &\mbox{if}\ D_i=\pm 2,\\
d_i-D_i             &\mbox{if}\ D_i\in[-1,1]_\Z,
\end{array}\right.
S(D_i)=\left\{
\begin{array}{ll}
\emptyset           &\mbox{if}\ D_i=\pm 2,\\
\Si_i               &\mbox{if}\ D_i\in[-1,1]_\Z,
\end{array}\right.
$$
 $$  
\mb{and}\ \ S'(D_i)=\left\{
\begin{array}{ll}
\emptyset                           &\mbox{if}\ D_i=\pm 2,\\
 \{-1,0\}                           &\mbox{if}\ D_i=-1,\\
 \{-1,0,1\}             &\mbox{if}\ D_i=0,\\
 \{0,1\}              &\mbox{if}\ D_i=1.
\end{array}\right.
$$

Let $W=\{\ket{w_{p,\vec{\si_1},\vec{\si},\vec{\si_3}}}
|\ (p,\vec{\si_1},\vec{\si},\vec{\si_3})
\in Q\times\Si^3\}^{\bot\bot}$ and $V=
\{\ket{v_{p,\vec{\si_1},\vec{\si},\vec{\si_3}}}
|\ (p,\vec{\si_1},\vec{\si},\vec{\si_3})
\in Q\times\Si^3\}^{\bot\bot}$. 
 By the unitarity conditions of the quantum transition functions
 of multi-tape QTM \cite{ON98}, the subspaces $W$, $V$ and $H$
 are all orthogonal one another
 and it is verified that
 $\{\ket{v_{p,\vec{\si_1},\vec{\si},\vec{\si_3}}}\}$
 is an orthonormal system of $V$. Thus, there exists a quantum gate $G_1$
 satisfying the above condition. 
 The subcircuit $K$ simulating one step of $M$ consists of
 $(2t-1)^k$ quantum gates $G_1$ and a reversible Boolean gate $G_2$,
 which works as the case of single tape QTMs,
 and for $i_1,\ldots,i_k=0,1,\ldots,2t-2$
 the $(\sum_{j=1}^{k}i_j(2t-1)^{k-j})$-th $G_1$ is connected
 with first $l_0$ wires and the wires of bit numbers
 $l_0+i_1l_1+1,\ldots,l_0+i_1l_1+l_1-1,
 l_0+(2t-1)l_1+i_2l_2+1,\ldots,l_0+(2t-1)l_1+i_2l_2+l_2-1,
 \ldots,l_0+(2t-1)(\sum_{j=1}^{k-1}l_j)+i_kl_k+1,\ldots,
 l_0+(2t-1)(\sum_{j=1}^{k-1}l_j)+i_kl_k+l_k-1$.
 Here, $l_j=2+\lceil \log|\Si_j|\rceil$.
 In the case of $k=2$, the subcircuit $K$ is illustrated in Figure 6.
 Similar to the case of single tape QTMs, we can see
 that $t$ consecutive subcircuits $t$-simulates $M$.
 Therefore, Theorem \ref{th:43} holds for arbitrary $k$-tape QTMs.

\begin{center}\unitlength 3.5mm
\begin{picture}(30,27)

\put(2,5){\framebox(1.5,0.5)}\put(2,6){\framebox(1.5,0.5)}
\put(2,9){\framebox(1.5,0.5)}\put(2,10){\framebox(1.5,0.5)}
\put(4,5){\framebox(5.5,0.5)}\put(4,6){\framebox(5.5,0.5)}
\put(4,9){\framebox(5.5,0.5)}\put(4,10){\framebox(5.5,0.5)}
\put(16,5){\framebox(5.5,0.5)}\put(18,6){\framebox(5.5,0.5)}
\put(22,9){\framebox(5.5,0.5)}\put(24,10){\framebox(5.5,0.5)}

\put(2.5,4.5){\line(0,1){0.5}}\put(3,4.5){\line(0,1){0.5}}
\put(2.5,5.5){\line(0,1){0.5}}\put(3,5.5){\line(0,1){0.5}}
\put(2.5,6.5){\line(0,1){0.5}}\put(3,6.5){\line(0,1){0.5}}
\put(2.5,9.5){\line(0,1){0.5}}\put(3,9.5){\line(0,1){0.5}}
\put(2.5,10.5){\line(0,1){0.5}}\put(3,10.5){\line(0,1){0.5}}

\put(4.5,4.5){\line(0,1){0.5}}\put(5,4.5){\line(0,1){0.5}}
\put(4.5,5.5){\line(0,1){0.5}}\put(5,5.5){\line(0,1){0.5}}
\put(4.5,6.5){\line(0,1){0.5}}\put(5,6.5){\line(0,1){0.5}}
\put(4.5,9.5){\line(0,1){0.5}}\put(5,9.5){\line(0,1){0.5}}
\put(4.5,10.5){\line(0,1){15.5}}\put(5,10.5){\line(0,1){15.5}}

\put(6.5,4.5){\line(0,1){0.5}}\put(7,4.5){\line(0,1){0.5}}
\put(6.5,5.5){\line(0,1){0.5}}\put(7,5.5){\line(0,1){0.5}}
\put(6.5,6.5){\line(0,1){0.5}}\put(7,6.5){\line(0,1){0.5}}
\put(6.5,9.5){\line(0,1){0.5}}\put(7,9.5){\line(0,1){0.5}}
\put(6.5,10.5){\line(0,1){0.5}}\put(7,10.5){\line(0,1){0.5}}
\put(6.5,11.5){\line(0,1){0.5}}\put(7,11.5){\line(0,1){0.5}}
\put(6.5,12.5){\line(0,1){0.5}}\put(7,12.5){\line(0,1){0.5}}
\put(6.5,15.5){\line(0,1){0.5}}\put(7,15.5){\line(0,1){0.5}}
\put(6.5,16.5){\line(0,1){9.5}}
\put(7,16.5){\line(0,1){9.5}}

\put(8.5,4.5){\line(0,1){0.5}}\put(9,4.5){\line(0,1){0.5}}
\put(8.5,5.5){\line(0,1){0.5}}\put(9,5.5){\line(0,1){0.5}}
\put(8.5,6.5){\line(0,1){0.5}}\put(9,6.5){\line(0,1){0.5}}
\put(8.5,9.5){\line(0,1){0.5}}\put(9,9.5){\line(0,1){0.5}}
\put(8.5,10.5){\line(0,1){0.5}}\put(9,10.5){\line(0,1){0.5}}
\put(8.5,11.5){\line(0,1){0.5}}\put(9,11.5){\line(0,1){0.5}}
\put(8.5,12.5){\line(0,1){0.5}}\put(9,12.5){\line(0,1){0.5}}
\put(8.5,15.5){\line(0,1){0.5}}\put(9,15.5){\line(0,1){0.5}}
\put(8.5,16.5){\line(0,1){0.5}}
\put(9,16.5){\line(0,1){0.5}}

\put(10.5,4.5){\line(0,1){6.5}}\put(11,4.5){\line(0,1){6.5}}
\put(10.5,11.5){\line(0,1){0.5}}\put(11,11.5){\line(0,1){0.5}}
\put(10.5,12.5){\line(0,1){0.5}}\put(11,12.5){\line(0,1){0.5}}
\put(10.5,15.5){\line(0,1){0.5}}\put(11,15.5){\line(0,1){0.5}}
\put(10.5,16.5){\line(0,1){0.5}}
\put(11,16.5){\line(0,1){0.5}}

\put(16.5,4.5){\line(0,1){0.5}}\put(17,4.5){\line(0,1){0.5}}
\put(16.5,5.5){\line(0,1){5.5}}\put(17,5.5){\line(0,1){5.5}}
\put(16.5,11.5){\line(0,1){8.5}}
\put(17,11.5){\line(0,1){8.5}}
\put(16.5,20.5){\line(0,1){5.5}}
\put(17,20.5){\line(0,1){5.5}}

\put(18.5,4.5){\line(0,1){0.5}}\put(19,4.5){\line(0,1){0.5}}
\put(18.5,5.5){\line(0,1){0.5}}\put(19,5.5){\line(0,1){0.5}}
\put(18.5,6.5){\line(0,1){4.5}}
\put(19,6.5){\line(0,1){4.5}}

\put(20.5,4.5){\line(0,1){0.5}}\put(21,4.5){\line(0,1){0.5}}
\put(20.5,5.5){\line(0,1){0.5}}\put(21,5.5){\line(0,1){0.5}}

\put(22.5,4.5){\line(0,1){1.5}}\put(23,4.5){\line(0,1){1.5}}

\put(28.5,4.5){\line(0,1){5.5}}
\put(29,4.5){\line(0,1){5.5}}
\put(28.5,10.5){\line(0,1){5.5}}\put(29,10.5){\line(0,1){5.5}}
\put(28.5,16.5){\line(0,1){8.5}}\put(29,16.5){\line(0,1){8.5}}
\put(28.5,24.5){\line(0,1){0.5}}
\put(29,24.5){\line(0,1){0.5}}

\put(2,11){\framebox(1.5,0.5)}\put(2,12){\framebox(1.5,0.5)}
\put(2,15){\framebox(1.5,0.5)}\put(2,16){\framebox(1.5,0.5)}
\put(6,11){\framebox(5.5,0.5)}\put(6,12){\framebox(5.5,0.5)}
\put(6,15){\framebox(5.5,0.5)}\put(6,16){\framebox(5.5,0.5)}
\put(16,11){\framebox(5.5,0.5)}\put(18,12){\framebox(5.5,0.5)}
\put(22,15){\framebox(5.5,0.5)}\put(24,16){\framebox(5.5,0.5)}
\put(2.5,11.5){\line(0,1){0.5}}\put(3,11.5){\line(0,1){0.5}}
\put(2.5,12.5){\line(0,1){0.5}}\put(3,12.5){\line(0,1){0.5}}
\put(2.5,15.5){\line(0,1){0.5}}\put(3,15.5){\line(0,1){0.5}}
\put(2.5,16.5){\line(0,1){0.5}}\put(3,16.5){\line(0,1){0.5}}

\put(2,20){\framebox(1.5,0.5)}\put(2,21){\framebox(1.5,0.5)}
\put(2,24){\framebox(1.5,0.5)}\put(2,25){\framebox(1.5,0.5)}
\put(10,20){\framebox(5.5,0.5)}\put(10,21){\framebox(5.5,0.5)}
\put(10,24){\framebox(5.5,0.5)}\put(10,25){\framebox(5.5,0.5)}
\put(16,20){\framebox(5.5,0.5)}\put(18,21){\framebox(5.5,0.5)}
\put(22,24){\framebox(5.5,0.5)}\put(24,25){\framebox(5.5,0.5)}
\put(2.5,20.5){\line(0,1){0.5}}\put(3,20.5){\line(0,1){0.5}}
\put(2.5,21.5){\line(0,1){0.5}}\put(3,21.5){\line(0,1){0.5}}
\put(2.5,24.5){\line(0,1){0.5}}\put(3,24.5){\line(0,1){0.5}}
\put(2.5,25.5){\line(0,1){0.5}}\put(3,25.5){\line(0,1){0.5}}

\put(14.5,4.5){\line(0,1){15.5}}\put(15,4.5){\line(0,1){15.5}}
\put(14.5,20.5){\line(0,1){0.5}}\put(15,20.5){\line(0,1){0.5}}
\put(14.5,21.5){\line(0,1){0.5}}\put(15,21.5){\line(0,1){0.5}}
\put(14.5,24.5){\line(0,1){0.5}}\put(15,24.5){\line(0,1){0.5}}
\put(14.5,25.5){\line(0,1){0.5}}\put(15,25.5){\line(0,1){0.5}}

\put(10.5,21.5){\line(0,1){0.5}}\put(11,21.5){\line(0,1){0.5}}
\put(10.5,24.5){\line(0,1){0.5}}\put(11,24.5){\line(0,1){0.5}}
\put(10.5,25.5){\line(0,1){0.5}}\put(11,25.5){\line(0,1){0.5}}

\put(12.5,21.5){\line(0,1){0.5}}\put(13,21.5){\line(0,1){0.5}}
\put(12.5,24.5){\line(0,1){0.5}}\put(13,24.5){\line(0,1){0.5}}
\put(12.5,25.5){\line(0,1){0.5}}\put(13,25.5){\line(0,1){0.5}}

\put(18.5,11.5){\line(0,1){0.5}}\put(19,11.5){\line(0,1){0.5}}
\put(18.5,20.5){\line(0,1){0.5}}\put(19,20.5){\line(0,1){0.5}}
\put(18.5,21.5){\line(0,1){4.5}}\put(19,21.5){\line(0,1){4.5}}

\put(20.5,11.5){\line(0,1){0.5}}\put(21,11.5){\line(0,1){0.5}}
\put(20.5,20.5){\line(0,1){0.5}}\put(21,20.5){\line(0,1){0.5}}
\put(20.5,21.5){\line(0,1){0.5}}\put(21,21.5){\line(0,1){0.5}}

\put(22.5,11.5){\line(0,1){0.5}}\put(23,11.5){\line(0,1){0.5}}
\put(22.5,20.5){\line(0,1){0.5}}\put(23,20.5){\line(0,1){0.5}}
\put(22.5,21.5){\line(0,1){0.5}}\put(23,21.5){\line(0,1){0.5}}

\put(24.5,9.5){\line(0,1){0.5}}\put(25,9.5){\line(0,1){0.5}}
\put(24.5,15.5){\line(0,1){0.5}}\put(25,15.5){\line(0,1){0.5}}
\put(24.5,24.5){\line(0,1){0.5}}\put(25,24.5){\line(0,1){0.5}}
\put(24.5,25.5){\line(0,1){0.5}}\put(25,25.5){\line(0,1){0.5}}

\put(26.5,9.5){\line(0,1){0.5}}\put(27,9.5){\line(0,1){0.5}}
\put(26.5,15.5){\line(0,1){0.5}}\put(27,15.5){\line(0,1){0.5}}
\put(26.5,24.5){\line(0,1){0.5}}\put(27,24.5){\line(0,1){0.5}}
\put(26.5,25.5){\line(0,1){0.5}}\put(27,25.5){\line(0,1){0.5}}

\put(28.5,25.5){\line(0,1){0.5}}\put(29,25.5){\line(0,1){0.5}}

\put(3.5,5.25){\line(1,0){0.5}}\put(3.5,6.25){\line(1,0){0.5}}
\put(3.5,9.25){\line(1,0){0.5}}\put(3.5,10.25){\line(1,0){0.5}}
\put(3.5,11.25){\line(1,0){2.5}}\put(3.5,12.25){\line(1,0){2.5}}
\put(3.5,15.25){\line(1,0){2.5}}\put(3.5,16.25){\line(1,0){2.5}}
\put(3.5,20.25){\line(1,0){6.5}}\put(3.5,21.25){\line(1,0){6.5}}
\put(3.5,24.25){\line(1,0){6.5}}\put(3.5,25.25){\line(1,0){6.5}}

\put(9.5,5.25){\line(1,0){6.5}}\put(9.5,6.25){\line(1,0){8.5}}
\put(9.5,9.25){\line(1,0){12.5}}\put(9.5,10.25){\line(1,0){14.5}}
\put(11.5,11.25){\line(1,0){4.5}}\put(11.5,12.25){\line(1,0){6.5}}
\put(11.5,15.25){\line(1,0){10.5}}\put(11.5,16.25){\line(1,0){12.5}}
\put(15.5,20.25){\line(1,0){0.5}}\put(15.5,21.25){\line(1,0){2.5}}
\put(15.5,24.25){\line(1,0){6.5}}\put(15.5,25.25){\line(1,0){8.5}}

\put(2,26){\framebox(27.5,0.5)}

\put(0,2){cell}\put(2.5,2){P}
\put(4,2){$(-t,1)$}\put(13,2){$(t,1)$}\put(16,2){$(-t,2)$}\put(28,2){$(t,2)$}
\put(0,4.5){$G_1$}\put(0,6){$G_1$}\put(0,26){$G_2$}
\multiput(8.5,2.3)(1,0){4}{\line(1,0){0.1}}
\multiput(20.5,2.3)(1,0){7}{\line(1,0){0.1}}
\end{picture}

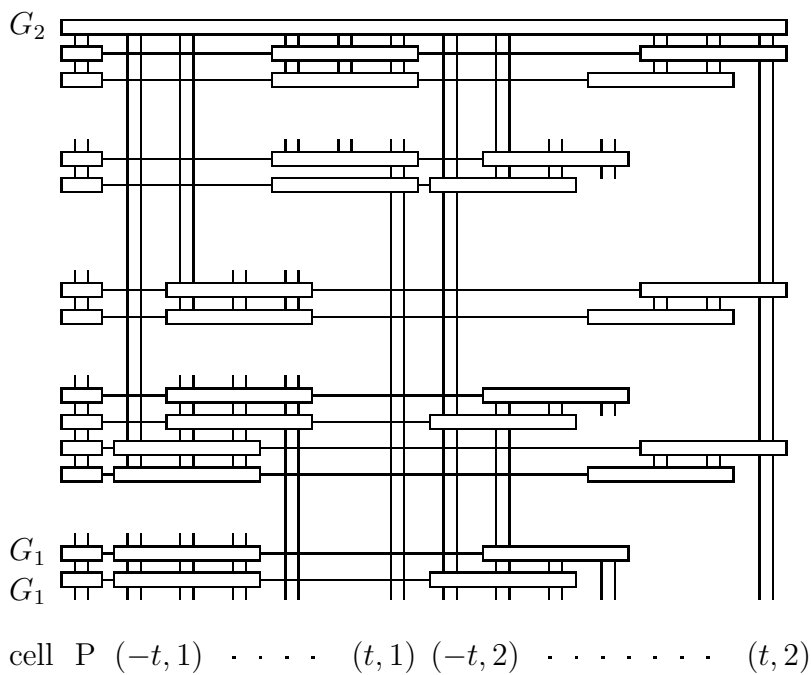
Figure 6 : The quantum circuit $K$ that simulates one step
 of a two-tape QTM $M$
\end{center}

\ \\ 
{\bf Acknowledgements}\ \

\ \

We thank John Watrous for helpful comments.
 H.N. thanks Tatsuie Tsukiji and Yasuo Yoshinobu
 for helpful discussions.

\end{document}